\documentclass[twocolumn, tighten,times]{aastex631} 
\usepackage{graphicx}
\usepackage{epstopdf}
\usepackage{amsmath}
\usepackage{booktabs}
\usepackage{tabularx}
\usepackage{float}
\usepackage{bm}
\usepackage{color}
\usepackage{soul}
\usepackage{xcolor}
\usepackage{multirow}

\DeclareUnicodeCharacter{02BC}{}

\shorttitle{JWST/CANDELS HALF MASS RADIUS}
\shortauthors{Xin. ET AL}
\graphicspath{{./}{figures/}}
\begin{document}

\title{Difference Between Half-mass Radius and Half-light Radius of Galaxies at 0.2 $< z <$ 2.5 Revealed by JWST/NIRCam Data}

\author[0009-0003-2773-0476]{Mingen Xin}
\affil{School of Mathematics and Physics, Anqing Normal University, Anqing 246133, China;\url{wen@mail.ustc.edu.cn}}
\affil{Institute of Astronomy and Astrophysics, Anqing Normal University, Anqing 246133, China}

\author[0000-0001-9694-2171]{Guanwen Fang}
\altaffiliation{Corresponding author: Guanwen Fang}
\affil{School of Mathematics and Physics, Anqing Normal University, Anqing 246133, China;\url{wen@mail.ustc.edu.cn}}
\affil{Institute of Astronomy and Astrophysics, Anqing Normal University, Anqing 246133, China}

\author[0000-0002-0846-7591]{Jie Song}
\affil{Department of Astronomy, University of Science and Technology of China, Hefei 230026, China;\url{jiesong@mail.ustc.edu.cn}} 
\affil{School of Astronomy and Space Science, University of Science and Technology of China, Hefei 230026, China}
\affil{Institute of Deep Space Sciences, Deep Space Exploration Laboratory, Hefei 230026, China;\url{xkong@ustc.edu.cn}}

\author[0000-0001-5988-2202]{Shiying Lu}
\affil{School of Mathematics and Physics, Anqing Normal University, Anqing 246133, China;\url{wen@mail.ustc.edu.cn}}
\affil{Institute of Astronomy and Astrophysics, Anqing Normal University, Anqing 246133,  China}
\affil{Key Laboratory of Modern Astronomy and Astrophysics (Nanjing University), Ministry of Education, Nanjing 210093, China}

\author[0000-0001-8078-3428]{Zesen Lin}
\affil{Institute for Astrophysics, School of Physics, Zhengzhou University, Zhengzhou, 450001, China}

\author[0000-0002-7660-2273]{Xu Kong}
\affil{Department of Astronomy, University of Science and Technology of China, Hefei 230026, China;\url{jiesong@mail.ustc.edu.cn}}
\affil{School of Astronomy and Space Science, University of Science and Technology of China, Hefei 230026, China}
\affil{Institute of Deep Space Sciences, Deep Space Exploration Laboratory, Hefei 230026, China;\url{xkong@ustc.edu.cn}}

\begin{abstract}
Using JWST observations in CANDELS fields, we measure the half-light radius ($r_{\rm e,light}$) and half-mass radius ($r_{\rm e,mass}$) for 14,333 galaxies with stellar masses $M_* > 10^9 M_\odot$ at redshifts $0.2 < z < 2.5$. To investigate the difference between $r_{\rm e,light}$ and $r_{\rm e,mass}$, we find that $r_{\rm e,light}$ is larger than $r_{\rm e,mass}$ for both quiescent galaxies (QGs) and star-forming galaxies (SFGs). Moreover, the difference between these two radius is clearly correlated with galaxy stellar mass, $r_{\rm e,light}$, and the rest-frame $U - V$ color. When examining the evolution of the $r_{\rm e,mass}/r_{\rm e,light}$ ratio, we observe a significant increase for SFGs at $z > 1.7$. In contrast, no clear increase is observed for QGs at $z > 1$, though a slight decreasing trend is seen between $0.2 < z < 1.0$. By fitting a linear relationship between galaxy size and stellar mass, we find that the slope for $r_{\rm e,light}$ is $\sim$ 0.1–0.3 dex larger than that for $r_{\rm e,mass}$. In terms of galaxy size evolution at a fixed stellar mass, the $r_{\rm e,mass}$ of QGs increases by a factor of $\sim$ 3–5 from $z \sim 2.5$ to $z \sim 0.2$. In contrast, the $r_{\rm e,mass}$ of SFGs increases by a factor of approximately 2 over the same redshift range, with this growth trend closely following that of their $r_{\rm e,light}$. These results indicate that previous insights into galaxy evolution based on $r_{\rm e,light}$ remain valid when considering $r_{\rm e,mass}$, although the specific slopes show some variations.

\end{abstract}

\section{Introduction} \label{sec:1}
The study of galaxy sizes and their evolution over cosmic time has been ongoing for decades and serves as a critical test for galaxy evolution models ({e.g., \citealt{Mo_1998,Kravtsov_2013,Jiang_2019}}). One of the key methods for understanding how stellar mass is assembled within a galaxy is the relationship between galaxy size and stellar mass (e.g., \citealt{vanderWel_2014,Lange_2015,Mosleh_2020,Miller_2023}). Numerous previous studies have examined this relation across a wide range of redshifts ($0<z<4$) and stellar masses ($10^{9.0}M_\odot <M_{\ast} < 10^{11.5}M_\odot$), revealing that both quiescent (QGs) and star-forming galaxies (SFGs) exhibit a positive correlation in their mass-size relation (e.g., \citealt{Shen_2003, Ferguson_2004, Trujillo_2006, Williams_2010, vanderWel_2014, Mowla_2019,vanderWel_2024,ji2024jadesrestframeuvtonirsize,Jia_2024,Martorano_2024}). Moreover, the slopes of the mass-size relation for QGs are generally steeper compared to SFGs (e.g., \citealt{vanderWel_2014, Mowla_2019}). Additionally, several studies have also shown that, at a fixed stellar mass, galaxies at lower redshift ($0<z<2$) exhibit larger sizes compared to those at higher redshift ($z>2$), which is likely driven by a 
complication of the ``inside-out'' growth mechanism and the ``progenitor bias'' effect. (e.g., \citealt{vanDokkum_2001,Bezanson_2009, Naab_2009, Carollo_2013, Poggianti_2013,vanDeSande_2013}).

However, these studies are primarily based on the light-weighted morphologies (particularly in the optical band) of galaxies. Since most galaxies exhibit negative radial color gradients and, consequently, negative mass-to-light ratio ($M/L$) gradients, light serves as a biased tracer of stellar mass (e.g., \citealt{Szomoru_2013,Tacchella_2016,Suess_2019, Miller_2023,Buitrago_2024,Chamba_2024}). Such negative color gradients may be attributed to a variety of mechanisms. For QGs, the color gradient is often attributed to lower metallicity in the outer regions, while for SFGs, it is typically due to the presence of distinct components: an older, redder bulge at the center and a younger, bluer disk at larger radii (e.g., \citealt{Wu_2005, LaBarbera_2009, Miller_2023}). Therefore, the commonly used half-light radius ($r_{\rm e,light}$) does not directly probe the underlying stellar mass distribution of galaxies (e.g., \citealt{Mosleh_2017, Suess_2019, Miller_2023}). 

A more robust parameter for representing the size of a galaxy is its half-mass radius ($r_{\rm e,mass}$), which can be obtained by fitting the galaxy's stellar mass map using a S\'{e}rsic function. Several methods are available to obtain a galaxy's stellar mass map. Some studies derive the stellar mass map through the relationship between color and $M/L$ (e.g., \citealt{Szomoru_2013, Chan_2016, Suess_2019, vanderWel_2024}). However, given the significant scatter in the relationship between color and $M/L$, the stellar mass maps obtained through this method may be unreliable. Some other research had utilized the pixel-by-pixel spectral energy distribution (SED) fitting method to obtain the stellar mass map (e.g., \citealt{Suess_2019, Mosleh_2020}). This method can reliably obtain the distribution of various physical properties within galaxies and has been applied in many studies, including star formation rate profiles, color gradients, and galaxy mergers (e.g., \citealt{Lanyon-Foster_2007, Lanyon-Foster_2012, Zibetti_2009, Hemmati_2014, Cibinel_2019}). Thus, this method can be employed to investigate the $r_{\rm e,mass}$ of galaxies.

However, some recent studies have indicated that applying this method at high redshifts (for example, at $z>1$) may involve certain risks (e.g., \citealt{Zibetti_2009,Mart_2017,Sorba_2018,Papovich_2023,Song_2023,Williams_2024,Wang_2025}). To perform the pixel-by-pixel SED fitting, imaging with high resolution is crucial. Before the successful launch of the James Webb Space Telescope (JWST), all available data were derived from the Hubble Space Telescope (HST, e.g., \citealt{Jafariyazani_2019,Suess_2019, Mosleh_2020,Moein_2025}). But, it is important to note that the reddest band of the HST is F160W, which can only trace rest-frame wavelengths shorter than 1 $\mu m$ at $z>1$. Recent work by \cite{Song_2023} has shown that when using SED fitting, rest-frame near-infrared data plays a crucial role in accurately measuring galaxy stellar masses. When lacking data whose wavelength is longer than 1 $\mu m$ in the rest-frame, the stellar mass of galaxies may be overestimated by 0.2 dex. Therefore, the results about the evolution of $r_{\rm e,mass}$ derived solely from HST data require further validation.

Fortunately, the high-resolution near-infrared data provided by JWST offer a valuable opportunity for conducting more in-depth investigations about $r_{\rm e,mass}$. The release of JWST/NIRCam imaging over fields previously observed by CANDELS (e.g., GOODS, COSMOS, EGS, and UDS; \citealt{Giavalisco_2004, Scoville_2007, Davis_2007, Lawrence_2007, Grogin_2011, Koekemoer_2011}) enables more precise stellar mass mapping at $z > 1$.  In this study, we combine JWST/NIRCam data with existing HST observations to construct improved stellar mass maps for galaxies with $M_{\ast} > 10^9~M_{\odot}$ across $0.2 < z < 2.5$ using pixel-by-pixel SED fitting. We then measure both the $r_{\rm e,light}$ and $r_{\rm e,mass}$ for these galaxies.

We analyze the relationship between galaxy size and stellar mass using both $r_{\rm e,light}$ and $r_{\rm e,mass}$. Our results confirm positive correlations in both cases, though the slope of the $r_{\rm e,mass}$–$M_{\ast}$ relation is shallower than that of the $r_{\rm e,light}$–$M_{\ast}$ relation. We also examine how galaxy size evolves with redshift. While both $r_{\rm e,light}$ and $r_{\rm e,mass}$ increase over time, the degree of size growth is somewhat similar for $r_{\rm e,mass}$ and for $r_{\rm e,light}$, consistent with the previous JWST-based studies (e.g., \citealt{vanderWel_2024,Clausen_2025,ji2024jadesrestframeuvtonirsize,Jia_2024,Martorano_2024}).

This paper is organized as follows. Section \ref{sec:2} outlines the data and selection criteria used in this study. In Section \ref{sec:3}, we provide a brief overview of the methodology employed to measure $r_{\rm e, mass}$. Our results are presented in Section \ref{sec:4}, which is followed by a discussion in Section \ref{sec:5}. Finally, the conclusions are summarized in Section \ref{sec:6}. Throughout the paper, we adopt a flat $\Lambda$CDM cosmology with $H_0 = 70~\mathrm{km~s^{-1}~Mpc^{-1}}$, $\Omega_{\rm{m}}=0.3$, and $\Omega_{\rm{\Lambda}}=0.7$, and a \cite{Chabrier_2003} initial mass function (IMF).

 \begin{figure}

 \includegraphics[width=1\linewidth]{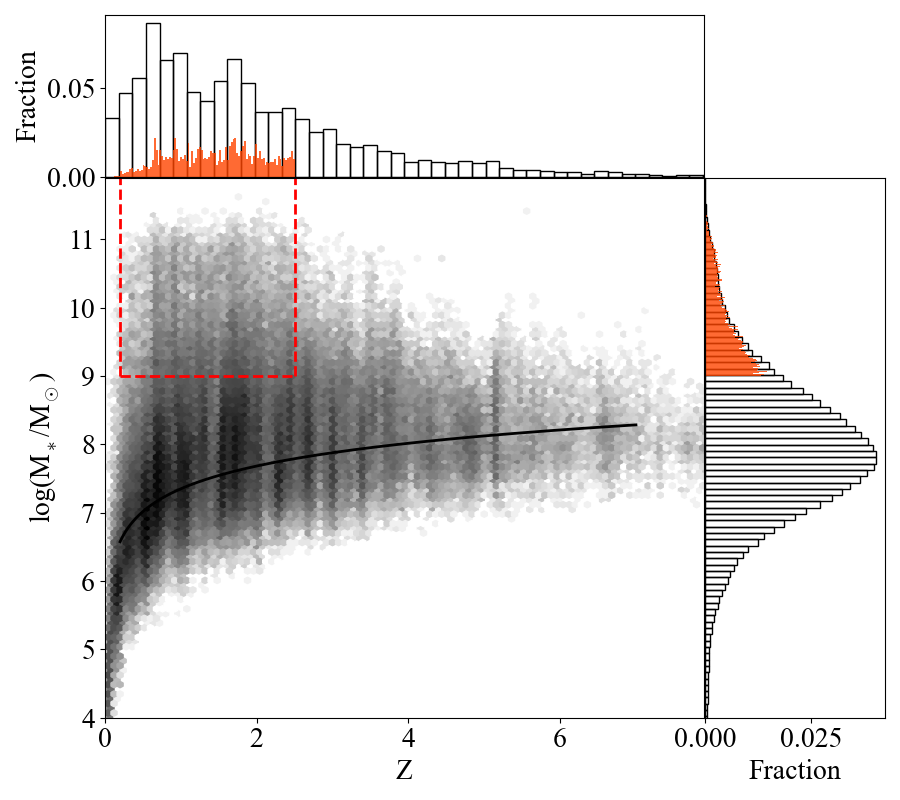}
 \caption {The redshift and stellar mass distribution of our total sample. The region enclosed by the red dashed lines indicates the selected sample used in this study, defined by $0.2<z<2.5$ and $\log(M_{\ast}/M_{\odot}) > 9$.  The black solid curve represents the 90\% stellar mass completeness limit corresponding to a magnitude limit of $\rm F444W_{lim}$ = 28.0 mag. The gray (red) histograms in the top and right panels show the redshift and stellar mass distributions of the total sample (selected sample), respectively.}
\label{fig:1}
\end{figure}  

\section{Data and Sample Selection} \label{sec:2}

\subsection{Dataset} \label{sec:2.1}
In this work, we utilize imaging data from three major JWST extragalactic surveys: the Cosmic Evolution Early Release Science Survey (CEERS; \citealt{Finkelstein_2023}), the JWST Advanced Deep Extragalactic Survey (JADES, including JADES-GDS and JADES-GDN; \citealt{Eisenstein_2023}), and the Public Release IMaging for Extragalactic Research (PRIMER, including PRIMER-COSMOS and PRIMER-UDS; \citealt{2021jwst.prop.1837D}). The CEERS program covers an area of $94.6~\text{arcmin}^2$ within the EGS field. We make use of six NIRCam broad-band filters available in this field: F115W, F150W, F200W, F277W, F356W, and F444W. The JADES survey spans approximately $177~\text{arcmin}^2$ across the GOODS-N and GOODS-S regions, while PRIMER covers roughly $393~\text{arcmin}^2$ in the COSMOS and UDS fields. For both JADES and PRIMER, we include observations in the same six bands as CEERS, along with F090W imaging. Although medium-band NIRCam observations are available in portions of these fields, they are significantly shallower than the broad-band data.  
Thus, we restrict our analysis to the broad-band images. Moreover, we have tested that the lack of F090W data in the CEERS field does not have a significant impact on our results, so we include the CEERS field in our subsequent analyses. To supplement the JWST data at shorter wavelengths, we incorporate HST imaging in the optical bands F435W, F606W, and F814W. This extended wavelength baseline improves the reliability of photometric redshift and stellar population estimates.

All JWST and HST imaging used in this work have been uniformly processed by \citet{Valentino_2023}, and the reduced images are publicly available via the Cosmic Dawn Center\footnote{https://dawn-cph.github.io/dja/index.html}. In addition to imaging products, \citet{Valentino_2023} also provides source catalogs generated via multi-band detection, including photometry and photometric redshift estimates. In deriving photometric redshifts, \citet{Valentino_2023} have already taken into account the use of medium bands, since many studies have demonstrated that incorporating medium bands plays an important role in accurately determining galaxy redshifts. We also tested the effect of medium-band data on stellar mass estimates from SED fitting at fixed redshift, and found that the stellar masses derived with and without medium-band data are nearly identical.

Using the photometric catalogs and $z_{\rm phot}$ provided by the Cosmic Dawn Center, Song et al. (in prep) performed SED fitting with {\tt\string CIGALE} \citep{2022zndo...6825092B} to derive physical properties of galaxies in these fields. Their fitting procedure assumes a delayed-$\tau$ star formation history (SFR$(t) \propto t\exp(-t/\tau)$), the \citet{Bruzual&Charlot_2003} stellar population models, nebular emission following \citet{Inoue_2011}, and the dust attenuation law from \citet{Charlot_2000}. The parameter settings used in their modeling are consistent with those adopted in \cite{Shen_2024}. The resulting catalogs provide estimates of stellar mass and other physical properties, which we adopt in this work.

\subsection{Sample Selection} \label{sec:2.2}
To ensure the robustness of our analysis, we adopt a set of selection criteria following the recommendations of \cite{Valentino_2023} and Song et al. (in prep). Specifically, we select galaxies that satisfy the following conditions: (1) signal-to-noise ratio (S/N) greater than 10 in the detection image ($\text{S/N}_{\text{det}} > 10$) to ensure reliable source selection; (2) reduced chi-squared value ($\chi^2/N_{\text{fit}} \leq 8$) with at least six filters available for photometric redshift estimation ($N_{\text{fit}} \geq 6$) to ensure that $z_{\rm phot}$ is reliable; (3) S/N greater than 3 in all six JWST broad bands (F115W, F150W, F200W, F277W, F356W, F444W), which ensures reliable SED fitting; (4) $\log(M_{\ast} / M_{\odot}) > 9$; and (5) $0.2 < z < 2.5$. 

Since our analysis incorporates observations from multiple fields with varying depths, it is important to ensure that our sample is mass-complete across all fields. Using source injection simulations, \cite{Merlin_2024} demonstrate that the detection completeness reaches nearly 100\% in all fields for sources brighter than 28 mag. Based on $\rm F444W_{lim} = 28 mag$, Song et al. (in prep) derived the 90\% stellar mass completeness as a function of redshift using the following relation: $M_{\rm comp}(z) = 7.36 + 0.5 \ln(z)$. The stellar masses of galaxies in our sample lie more than 1 dex above the completeness threshold across the full redshift range. Additionally, we examine the F444W magnitude distribution of our sample and find that all galaxies are brighter than 26 mag. Therefore, we think that the varying observational depths across different fields do not affect our results. After applying the selection criteria, we obtain a final sample of 14,333 galaxies. The distributions of stellar mass and redshift are shown in the right and top panels of Figure~\ref{fig:1}, respectively. The joint distribution of stellar mass versus redshift is displayed in the center panel of Figure~\ref{fig:1}.

  \begin{figure*}

 \includegraphics[width=1\linewidth]{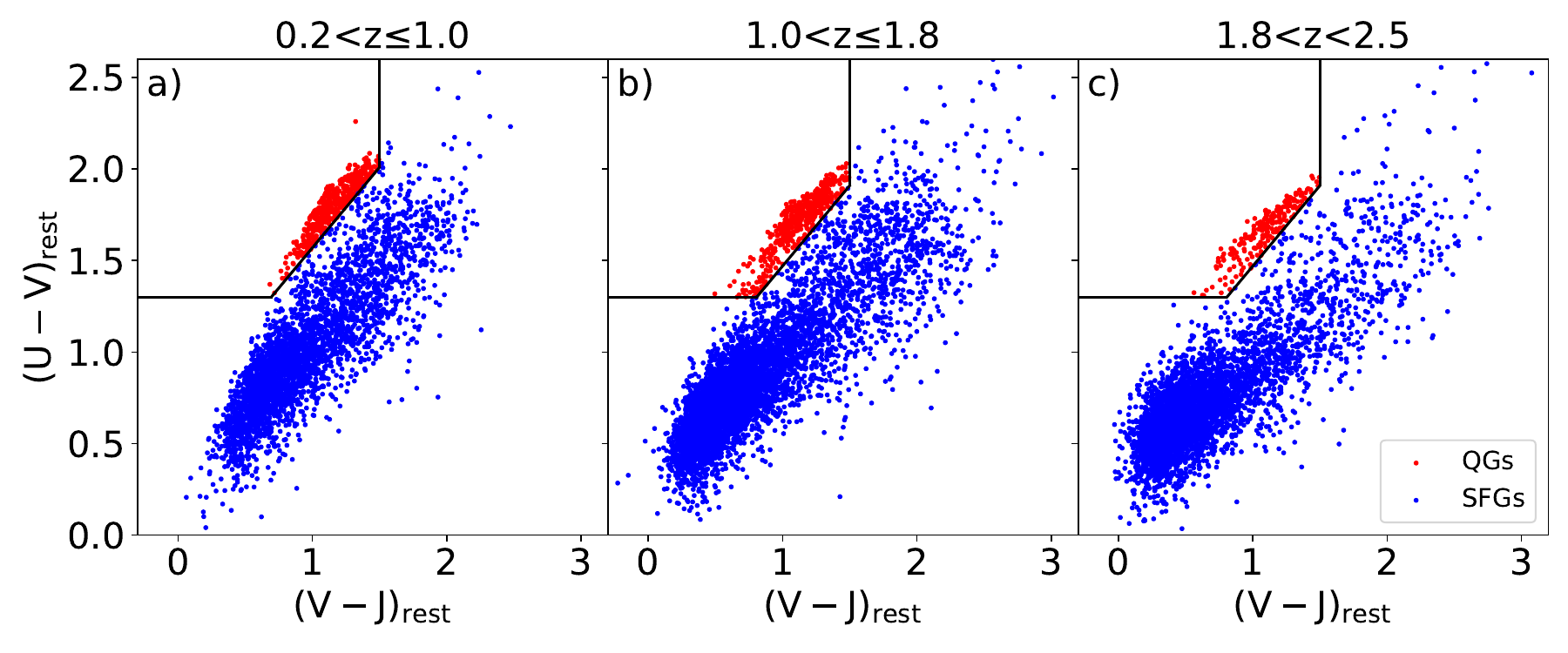}
 \caption {Rest-frame $U-V$ versus $V-J$ color distribution for three redshift bins. QGs and SFGs are separated by the selection criteria defined in Section \ref{sec:2.3}, shown by the black lines. SFGs are represented by blue points, while QGs are represented by red points. }
\label{fig:2}
\end{figure*}   
\subsection{Classifying Star-forming and Quiescent Galaxies} \label{sec:2.3}
To differentiate the results between SFGs and QGs, we first classify our sample into SFGs and QGs using the rest-frame $U - V$ and $V -J$ colors. As demonstrated by \cite{Franx_2003} and \cite{Whitaker_2012}, galaxies occupy different regions in the $UVJ$ plane, depending on their particular star formation rate and dust. In this paper, we adopt the criteria from \cite{Muzzin_2013} to distinguish SFGs from QGs, where QGs are defined as: 

\begin{align}
(U-V)_{\rm rest}  &> 1.3 && \text{for all } z \nonumber\\
(V - J)_{\rm rest} & < 1.5 && \text{for all } z \nonumber\\
(U-V)_{\rm rest} &> 0.88(V - J)_{\rm rest} + 0.69 && [0.0 < z < 1.0] \nonumber\\
(U-V)_{\rm rest} &> 0.88(V - J)_{\rm rest} + 0.59 && [1.0 < z < 3.0] \nonumber
\end{align}

\noindent
The distribution of our sample in the rest-frame $UVJ$ color space is shown in Figure \ref{fig:2}. From this figure, it is evident that the proportion of QGs in our sample is quite low, particularly at $z>1.8$. This is also consistent with galaxy evolution models, which predict that the proportion of QGs gradually increases as redshift decreases (\citealt{Marchesini_2014,Mowla_2019}). Finally, we obtained a sample of 13080 SFGs and 1253 QGs.

\begin{figure*}

 \includegraphics[width=1\linewidth]{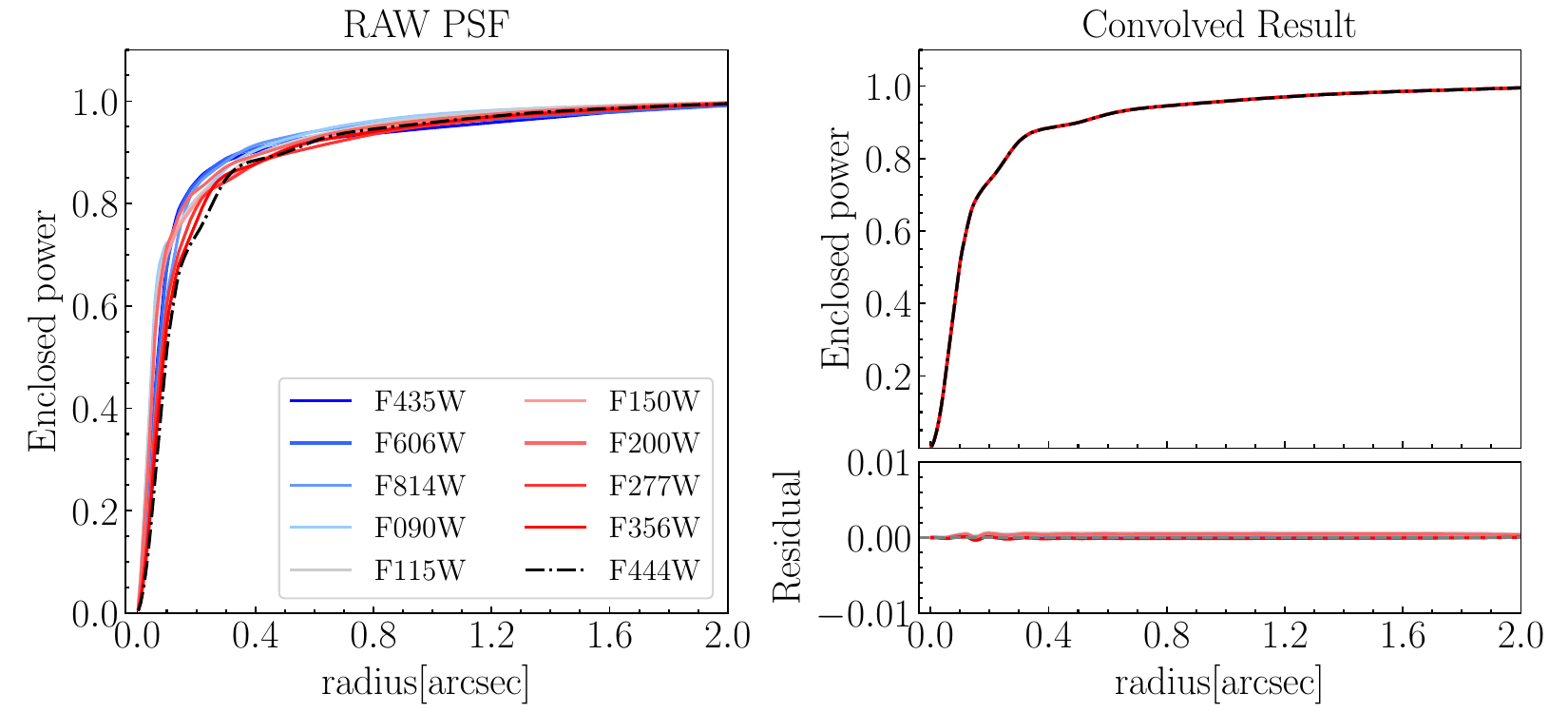}
 \caption {Left panel: Encircled energy profiles of the empirical PSFs in different HST and JWST filters measured in the JADES-GDS field. Right panel: Encircled energy profiles of the PSFs after convolution with the kernels, along with the residuals relative to the F444W PSF. The small residuals indicate that the PSFs have been successfully homogenized to match the F444W resolution.}
\label{fig:3}
\end{figure*}      

\begin{figure*}
 \includegraphics[width=1\linewidth]{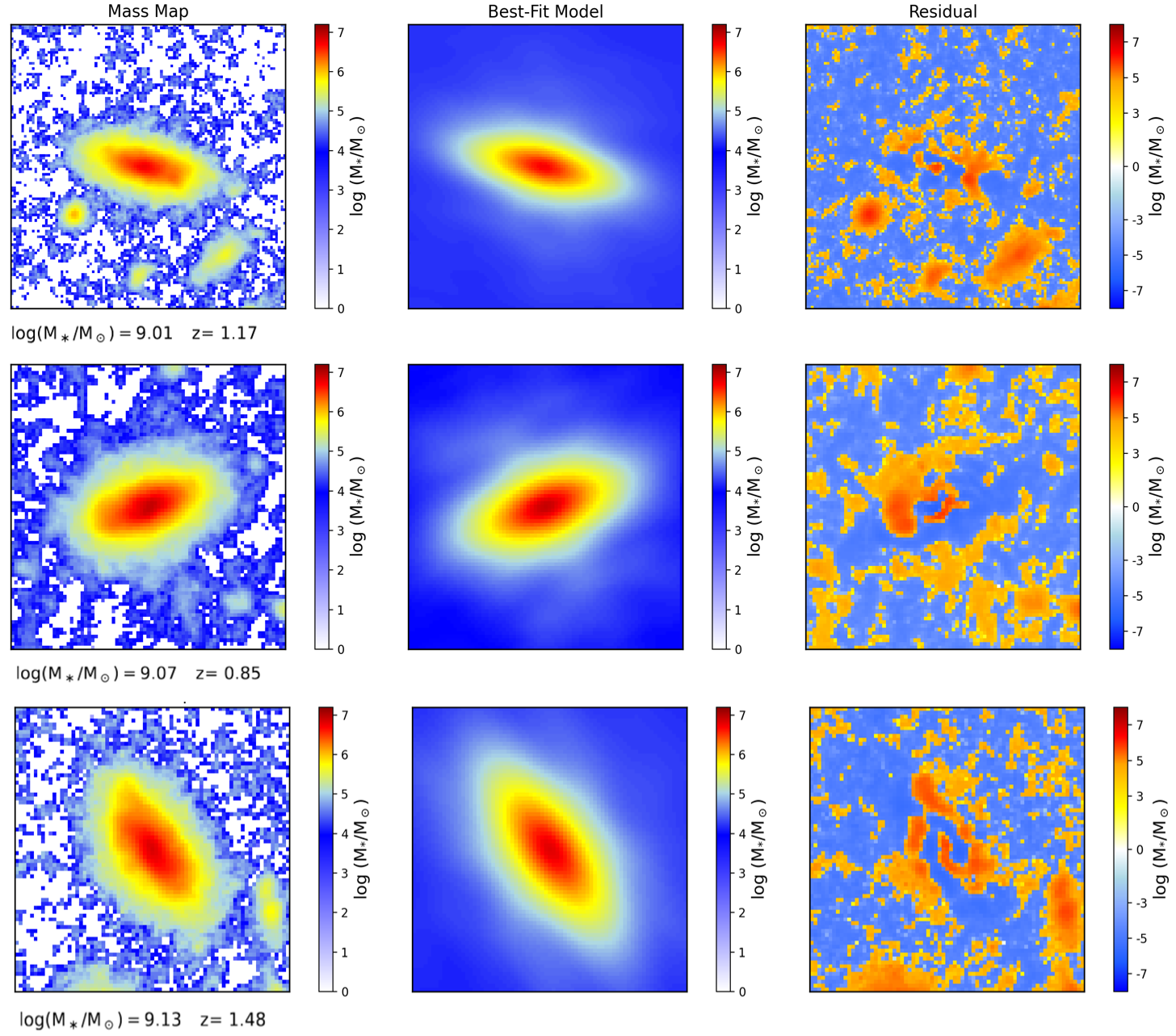} 
 \caption {Examples of stellar mass maps. The stellar mass maps (left panels) are derived using the pixel-by-pixel SED fitting method (Section~\ref{sec:3.3.1}). The best-fit models (middle panels) are obtained by fitting the stellar mass maps with \texttt{GALFIT}. The residuals (displayed in the right-hand panel) are obtained by subtracting the best-fit model from the stellar mass map. The redshift and stellar mass of these galaxies are listed at the bottom of each panel in the first column. }
\label{fig:4}
\end{figure*}    

\section{Methods} \label{sec:3}

\subsection{Image Processing}
To perform pixel-by-pixel SED fitting, it is essential to apply point-spread function (PSF) matching to ensure that the images across different bands have the same resolution. Therefore, we convolve all images to match the resolution of F444W in each field, which has the largest full width at half maximum (FWHM). The PSFs and corresponding convolution kernels used in this study were prepared by Song et al. (in prep). We summarize the procedure here. Numerous studies have shown that point sources form a tight sequence in the size–magnitude diagram (e.g., \citealt{Skelton_2014,Zhang_2024}), which allows for robust selection of stars for PSF construction. Following this method, we select 10–20 unsaturated point sources in each field. Using the \texttt{photutils} package \citep{2022zndo...6825092B}, we extract and combine the PSFs of the selected stars into a single, high-S/N PSF for each band using the \texttt{photutils.psf.EPSFBuilder} function, which recenters and normalizes each star before stacking. We compare our constructed PSFs with those provided by the Cosmic Dawn Center and found excellent agreement, confirming the reliability of our PSF measurements.

Convolution kernels for PSF matching are also derived using \texttt{photutils}. The reliability of the PSF matching is evaluated using the $D$ and $W_{-}$ metrics defined in \citet{Aniano_2011}, which characterize the fidelity of kernel performance. By iteratively optimizing the parameters of the kernel generation process to minimize both $D$ and $W_{-}$, high-fidelity convolution kernels are obtained. As an example, in Figure \ref{fig:3}, we show the PSF matching results for the JADES-GDS field. It can be seen that, after convolution with the kernels, the PSFs in different bands exhibit very good consistency with the F444W PSF, with deviations of less than 0.01, demonstrating the accuracy of our PSF matching. Then all images are convolved with the corresponding kernels to match the resolution of the F444W band, ensuring uniform spatial resolution across all bands used in the SED fitting. In addition to PSF homogenization, we apply corrections for Galactic extinction. The extinction values are estimated using the dust maps from \citet{Schlafly_2011}, assuming the \citet{Fitzpatrick_1999} extinction law with $\rm R_V = 3.1$.

\subsection{Half-light Radius} \label{sec:3.2}
For each galaxy, we perform \texttt{GALFIT} \citep{Peng_2010} fitting in every band to measure the $r_{\rm e,light}$, using the PSF-matched images and adopting the PSF from the F444W band. During the fitting, the effective radius is constrained to lie between 1 and 200 pixels, and the S\'{e}rsic index is limited to $n\sim0-8$. Then, we estimate the morphology at rest-frame 5000 \AA\ by
linearly interpolating the S\'{e}rsic parameters from the two observed bands that are nearest in wavelength. If only one suitable band is available, we adopt its fitted parameters directly. Although the observational depths vary somewhat across different fields, our sample is at least 2 mag brighter than the detection limit. Therefore, the variation in depth does not significantly affect our results.

\subsection{Half-mass Radius} \label{sec:3.3}
In this work, we adopt two different methods to measure the galaxy's $r_{\rm e,mass}$. The first method (2D stellar mass map method) is to derive the stellar mass map of each galaxy through pixel-by-pixel SED fitting, and then fit the 2D stellar mass map with a single S\'{e}rsic model using \texttt{GALFIT} \citep{Peng_2010} to obtain $r_{\rm e,mass}$, along with the corresponding uncertainties provided by GALFIT. However, some studies have shown that \texttt{GALFIT} tends to underestimate the uncertainties of morphological parameters (e.g., \citealt{Haussler_2007,Dimauro_2018}). In Section \ref{sec:5}, we evaluate the reliability of our method using a mock recovery test. Therefore, we use the results of this test to re-estimate the uncertainties in $r_{\mathrm{e, mass}}$. In brief, we divide the mock galaxies into several stellar-mass bins and, for each bin, compute the scatter of the difference between the recovered $r_{\mathrm{e, mass}}$ and the input values. Then, using the median stellar mass of each mass bin together with the corresponding uncertainties, we determine the $r_{\mathrm{e, mass}}$ uncertainty for each galaxy in our sample by interpolating. The second method (1D stellar mass density profile method) is to perform SED fitting on the multi-band fluxes measured within elliptical apertures of different radii, thereby constructing the 1D stellar mass density profile of the galaxy. We then fit this stellar mass density profile to derive $r_{\rm e,mass}$.

\subsubsection{Method 1: 2D Stellar Mass Map Method} \label{sec:3.3.1}
We construct stellar mass maps of the galaxies through pixel-by-pixel SED fitting and measure their $r_{\rm e, mass}$ using \texttt{GALFIT} \citep{Peng_2010}. The specific processing steps are described below: we first extract a $400 \times 400$ pixel cutout for each galaxy using the PSF-matched images. Then, we perform SED fitting for each pixel within the cutout using the \texttt{CIGALE} program (\citealt{Boquien_2019}).
When performing SED fitting, the redshift of each pixel is fixed to the galaxy's redshift, which is obtained from the Cosmic Dawn Center catalog, and the parameter settings are the same as described in Section \ref{sec:2.1}. Some examples of the stellar mass maps we obtained are shown in the left column of Figure \ref{fig:4}. By masking out other galaxies in the cutouts, we then measure $r_{\rm e, mass}$ of these galaxies using \texttt{GALFIT} with the same setup as described in Section \ref{sec:3.2}, which has been shown to reliably recover $r_{\rm e, mass}$ from mock data by \cite{Mosleh_2020}. The fitting results of stellar mass maps in Figure \ref{fig:4} are shown in the second column, with the corresponding residuals displayed in the third column. Figure~\ref{fig:4} reveals that the S\'{e}rsic model provides a good fit to the stellar mass map of the galaxy.

Since we perform SED fitting on each pixel of the galaxy, this approach may introduce potential issues, such as the neglect of correlated noise and the low S/N ratio in the outer regions, which could lead to unreliable fitting results. Some studies have suggested that pixel binning can mitigate these effects to some extent (e.g., \citealt{Abdurro'uf_2023,Novan_2025}). However, it is important to note that binning essentially acts as a smoothing process, which may partially wash out the intrinsic structural information of galaxies. Since different pixels are not completely independent, their fluxes can still partially reflect the variations within different regions of a galaxy. Simple pixel binning would smooth over these variations and reduce the spatial detail captured. Our goal is therefore to preserve as much of JWST’s high-resolution information as possible, motivating the choice of pixel-by-pixel SED fitting without binning. In Section \ref{sec:5}, we further demonstrate the reliability of our method for deriving $r_{\rm e, mass}$ using two complementary approaches: first, through mock simulation tests, and second, by comparing our results with those from previous studies.

It is worth noting that \texttt{GALFIT} \citep{Peng_2010} fitting fails for approximately 27.66\% of the galaxies. To assess the potential impact of these failures, we compared the distributions of stellar mass and other relevant properties between galaxies with successful and failed fits. We find that the distributions are similar, suggesting that these failures are unlikely to introduce significant biases in our results. Nevertheless, we caution that some residual effects cannot be completely ruled out, and readers should keep this limitation in mind when interpreting our measurements.

\begin{figure*}

 \includegraphics[width=1\linewidth]{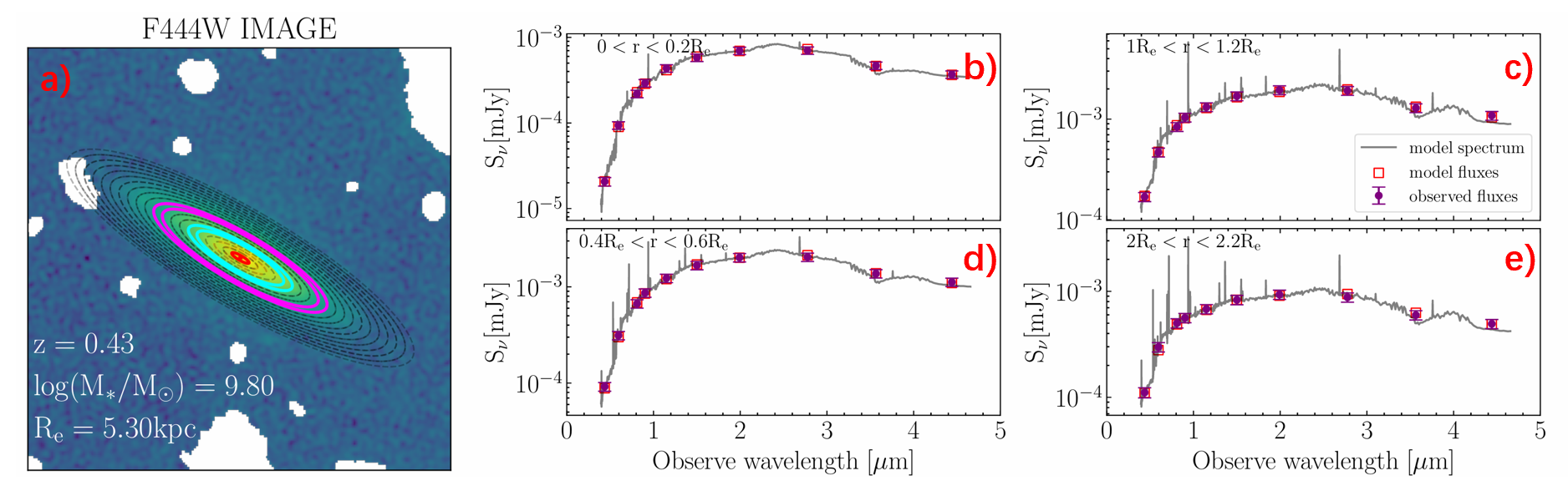}
 \caption {Schematic illustration of the method used to derive $r_{\rm e, mass}$ from the one-dimensional stellar mass surface density profile, and the comparison with results obtained from other methods. Panel a): F444W-band image of an example galaxy, with black dashed lines marking elliptical annuli spaced at intervals of $0.2 R_{\rm e}$. Panel b)-e): Four representative radial regions are highlighted in red, orange, cyan, and magenta, corresponding to $0<r<0.2r_{\rm e,1\mu m}$, $0.4r_{\rm e,1\mu m}<r<0.6r_{\rm e,1\mu m}$, $1r_{\rm e,1\mu m}<r<1.2r_{\rm e,1\mu m}$, and $2 r_{\rm e,1\mu m}<r<2.2r_{\rm e,1\mu m}$, respectively.}
\label{fig:5}
\end{figure*} 

\begin{figure}
 \includegraphics[width=1\linewidth]{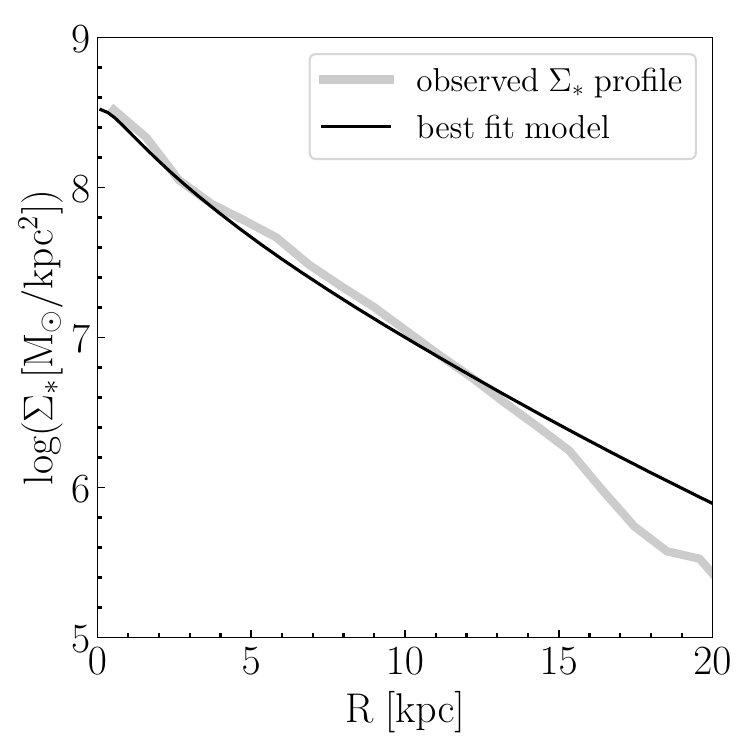}
 \caption {Example of a galaxy's 1D stellar mass surface density profile (gray line) along with the best-fit 1D S\'{e}rsic model (black line).}
\label{fig:6}
\end{figure}

\subsubsection{Method 2: 1D Stellar Mass Density Profile Method} \label{sec:3.3.2}

While the 2D stellar mass map method provides high-resolution measurements, it may be sensitive to pixel-level noise and outer low-S/N regions. To complement this, we also adopt an alternative approach based on 1D stellar mass density ($\Sigma_{\ast}$) profiles. For each galaxy, we can derive its $\Sigma_{\ast}$ profile by fitting spectral energy distributions at different radii. Then the galaxy's $r_{\rm e,mass}$ can be determined from these profiles. This method naturally accounts for correlated noise, mitigates the effects of low-S/N pixels in the outskirts, and is less sensitive to non-S\'{e}rsic structural components.

Specifically, we first estimate the galaxy morphologies at rest-frame $1\mu m$ to define the measurement apertures. Since we have already derived the galaxy morphologies in all available bands in section \ref{sec:3.2}, we can obtain the galaxy morphology at rest-frame $1\ \mu$m by linearly interpolating the S\'{e}rsic parameters from the observed bands. Then, we extract the fluxes in elliptical annuli with a step of $0.2 R_{\rm e}$ to obtain radial flux profiles, and perform resolved SED fitting in each elliptical annulus using {\ttfamily CIGALE}, adopting the same configurations described previously. An example of this procedure is shown in Figure \ref{fig:5}: panel a) presents a galaxy in the F444W band, with black dashed lines denoting elliptical annuli spaced at intervals of $0.2 R_{\rm e}$. The red, orange, cyan, and magenta annuli highlight four representative radial regions: $0<r<0.2r_{\rm e,1\mu m}$, $0.4r_{\rm e,1\mu m}<r<0.6r_{\rm e,1\mu m}$, $1r_{\rm e,1\mu m}<r<1.2r_{\rm e,1\mu m}$, and $2 r_{\rm e,1\mu m}<r<2.2r_{\rm e,1\mu m}$, respectively. Panels b)-e) present the corresponding SED fitting results for these four annuli. Using this method, we can obtain the $\Sigma_{\ast}$ profiles of galaxies.

Then, we fit the 1D $\Sigma_{\ast}$ profiles with a S\'{e}rsic function. Following \cite{Mosleh_2020}, we construct a library of S\'{e}rsic models to identify the best-fitting profiles. Specifically, we generate a grid of 15,400 S\'{e}rsic models spanning the $r_e - n$ parameter space (i.e., $0.01''<r_e<2''$ and $0.3<n<8$), with step sizes of $0.01''$ in $r_e$ and 0.1 in $n$. We then use \texttt{GALFIT} \citep{Peng_2010} to generate 2D model images by convolving the S\'{e}rsic models with the F444W-band PSFs, creating models separately for each field to account for field-to-field PSF variations. The 1D density profiles of the models are obtained by fitting ellipses to the 2D galaxy models (fixing the center, PA, and ellipticity). The best-fitting $\Sigma_{\ast}$ profiles are then determined by comparing the normalized 1D galaxy profiles with all normalized model profiles and finding the minimum $\chi^2$ value. Using this method, we can also obtain the $r_{\rm e, mass}$ of galaxies.

\begin{figure*}
 \includegraphics[width=1\linewidth]{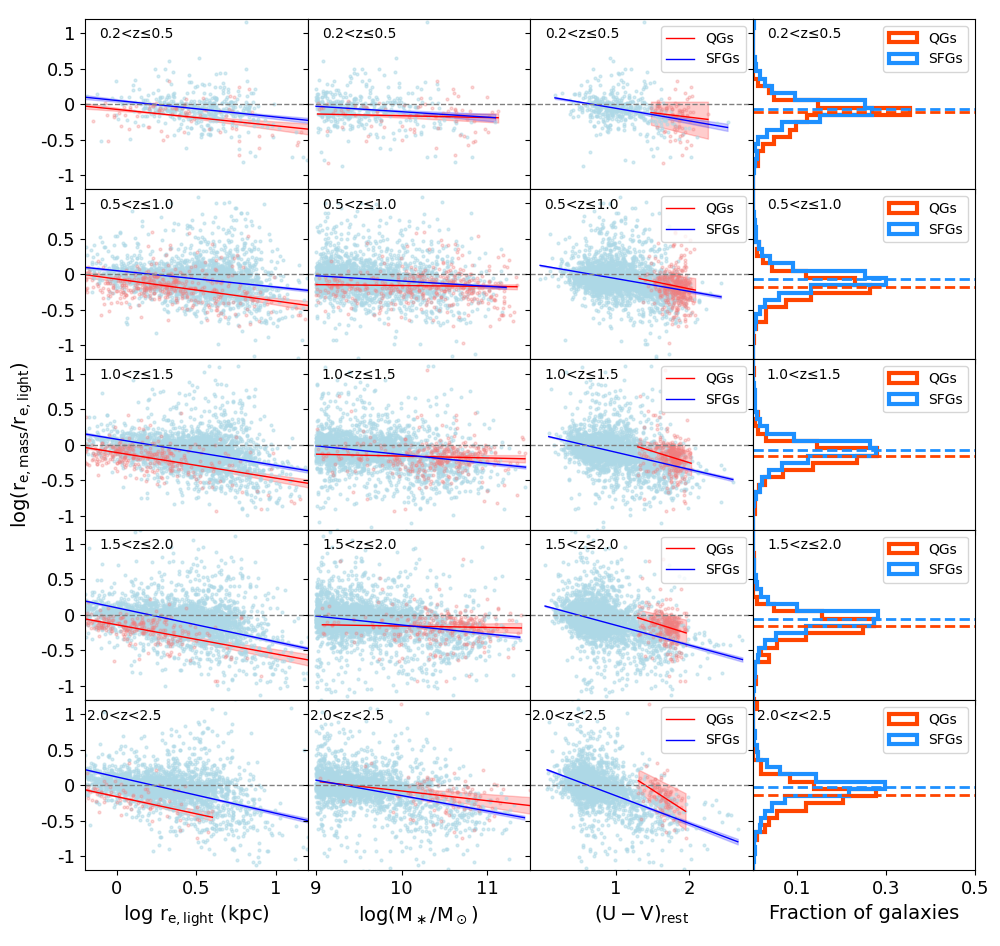}
 \caption {Relationships between the ratio of $r_{\rm e, mass}/ r_{\rm e, light}$ and different physical properties, including $r_{\rm e, light}$ (1st column), stellar mass (2nd column), (U-V)$_{\rm rest}$ color (3rd column), and the fraction of galaxies (4th column), respectively. From top to bottom, each row shows the corresponding relationship in different redshift bins. SFGs and QGs are displayed by light blue and light red dots, respectively. The best-fit result for SFGs (QGs) is also shown as the solid blue (red) line in each panel, which is calculated using a least-squares fit of all points. The shadow regions are determined by taking the 1$\sigma$ confidence intervals from 500 bootstrap samples. The horizontal centerline in the first three columns represents the 0 color gradient and t
he horizontal dashed lines in the 4th column indicates the median $r_{\rm e,mass} / r_{\rm e,light}$.}
\label{fig:7}
\end{figure*}

 \begin{figure}
 \includegraphics[width=1\columnwidth]{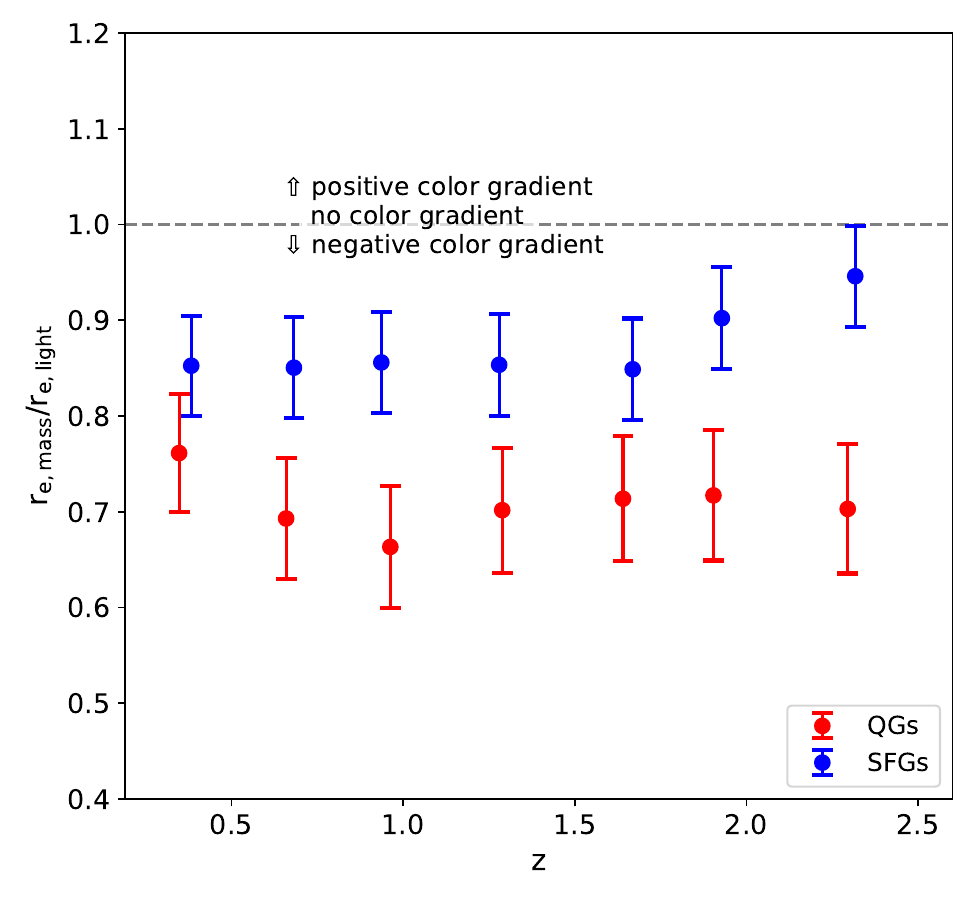}
\caption{
The median ratio of \(r_{\rm e,light} / r_{\rm e,mass}\) for SFGs (blue) and QGs (red) as a function of redshift. The corresponding error bars are estimated using 500 bootstrap resamplings, and the error for each galaxy is also accounted for by adding random perturbations.}
\label{fig:8}
\end{figure}

\section{Results } \label{sec:4}

In the previous section, we derive the stellar mass structural parameters of galaxies using both the 2D stellar mass map method and the 1D stellar mass density profile method. The following Section~\ref{5.1} will confirm that the results obtained from the two approaches are in good agreement. Therefore, in the subsequent analyses, we primarily adopt the results derived from Method~1. In this section, we explore the differences between these $r_{\rm e, mass}$ and $r_{\rm e, light}$ in detail and investigate the relationships between the galaxy size, stellar mass, and redshift of galaxies.

  \begin{table}
    \centering
    \caption{Best-fit parameters in different relationships.}
    \label{tab:1}
    \begin{tabular}{l|c|c|c}
  \hline\hline
        \multirow{2}{*}{Redshift} & Galaxy & \multirow{2}{*}{$s$} & \multirow{2}{*}{$b$} \\
         & Population &  &  \\
    \hline
        \multicolumn{4}{c}{$\log r_{\rm e,light}$ \;\; \; $\log r_{\rm {e,mass}}/r_{\rm {e,light}} = s(\log r_{\rm e,light}/\text{kpc} - 1) + b$}\\
  \hline
        $0.2 < z \leq 0.5$& SFGs & $-0.237^{+0.051}_{-0.039}$& $-0.171^{+0.026}_{-0.023}$\\
        $0.2  <  z \leq 0.5$& QGs & $-0.245^{+0.085}_{-0.065}$& $-0.316^{+0.059}_{-0.050}$\\
        $0.5  <  z \leq 1.0$& SFGs & $-0.227^{+0.022}_{-0.029}$& $-0.177^{+0.014}_{-0.017}$\\
        $0.5  <  z\leq 1.0$& QGs & $-0.312^{+0.067}_{-0.060}$& $-0.380^{+0.049}_{-0.042}$\\
        $1.0  <  z \leq 1.5$& SFGs & $-0.367^{+0.030}_{-0.031}$& $-0.292^{+0.020}_{-0.019}$\\
        $1.0  <  z\leq 1.5$& QGs & $-0.363^{+0.050}_{-0.048}$& $-0.472^{+0.046}_{-0.045}$\\
        $1.5 <  z  \leq 2.0$& SFGs & $-0.479^{+0.022}_{-0.026}$& $-0.379^{+0.016}_{-0.017}$\\
        $1.5 <  z  \leq   2.0$& QGs & $-0.413^{+0.072}_{-0.061}$& $-0.555^{+0.070}_{-0.057}$\\
        $2.0  <  z <2.5$& SFGs & $-0.513^{+0.030}_{-0.031}$& $-0.396^{+0.022}_{-0.023}$\\
        $2.0  <  z < 2.5$& QGs & $-0.479^{+0.101}_{-0.088}$& $-0.638^{+0.096}_{-0.088}$\\
\hline     
\multicolumn{4}{c}{$\log M_*/M_\odot$ \; \; \; $\log r_{\rm {e,mass}}/r_{\rm {e,light}} = s(\log M_\ast/M_\odot - 10) + b$}\\
\hline
        $0.2  <  z \leq 0.5$& SFGs & $-0.079^{+0.030}_{-0.031}$& $-0.184^{+0.026}_{-0.023}$\\
        $0.2  <  z \leq 0.5$& QGs & $-0.024^{+0.034}_{-0.032}$& $-0.167^{+0.023}_{-0.026}$\\
        $0.5  <  z \leq 1.0$& SFGs & $-0.074^{+0.012}_{-0.011}$& $-0.094^{+0.008}_{-0.007}$\\
        $0.5  <  z\leq 1.0$& QGs & $-0.013^{+0.018}_{-0.018}$& $-0.157^{+0.013}_{-0.011}$\\
        $1.0  <  z \leq 1.5$& SFGs & $-0.123^{+0.009}_{-0.009}$& $-0.141^{+0.007}_{-0.007}$\\
        $1.0  <  z\leq 1.5$& QGs & $-0.025^{+0.025}_{-0.026}$& $-0.162^{+0.015}_{-0.015}$\\
        $1.5  <  z \leq 2.0$& SFGs & $-0.126^{+0.012}_{-0.010}$& $-0.143^{+0.007}_{-0.007}$\\
        $1.5  <  z \leq2.0$& QGs & $-0.017^{+0.034}_{-0.032}$& $-0.160^{+0.023}_{-0.022}$\\
        $2.0  <  z <   2.5$& SFGs & $-0.215^{+0.013}_{-0.014}$& $-0.146^{+0.009}_{-0.009}$\\
        $2.0 <  z <  2.5$& QGs & $-0.136^{+0.051}_{-0.054}$& $-0.082^{+0.037}_{-0.032}$\\
\hline      
\multicolumn{4}{c}{ $(U-V)_{\rm rest}$ \; \; \; $\log r_{\rm {e,mass}}/r_{\rm {e,light}} = s((U-V)_{\rm rest} - 1) + b$}\\
\hline
        $0.2  <  z \leq 0.5$& SFGs & $-0.178^{+0.029}_{-0.027}$& $-0.060^{+0.009}_{-0.009}$\\
        $0.2  < z \leq 0.5$& QGs & $-0.145^{+0.133}_{-0.125}$& $-0.058^{+0.100}_{-0.098}$\\
        $0.5  <  z \leq 1.0$& SFGs & $-0.181^{+0.014}_{-0.011}$& $-0.059^{+0.008}_{-0.007}$\\
        $0.5  <  z\leq 1.0$& QGs & $-0.216^{+0.079}_{-0.091}$& $0.013^{+0.073}_{-0.064}$\\
        $1.0  <  z \leq 1.5$& SFGs & $-0.240^{+0.011}_{-0.012}$& $-0.107^{+0.004}_{-0.004}$\\
        $1.0  <  z\leq 1.5$& QGs & $-0.325^{+0.109}_{-0.087}$& $0.070^{+0.066}_{-0.080}$\\
        $1.5  <  z \leq 2.0$& SFGs & $-0.281^{+0.015}_{-0.013}$& $-0.146^{+0.006}_{-0.006}$\\
        $1.5  <  z \leq2.0$& QGs & $-0.324^{+0.121}_{-0.108}$& $-0.061^{+0.074}_{-0.087}$\\
        $2.0  <  z <   2.5$& SFGs & $-0.391^{+0.019}_{-0.019}$& $-0.148^{+0.008}_{-0.008}$\\
        $2.0 <  z <  2.5$& QGs & $-0.680^{+0.149}_{-0.162}$& $0.281^{+0.103}_{-0.101}$\\   
\hline
\end{tabular}
\end{table}

\begin{figure*}
 \includegraphics [width=1\linewidth]{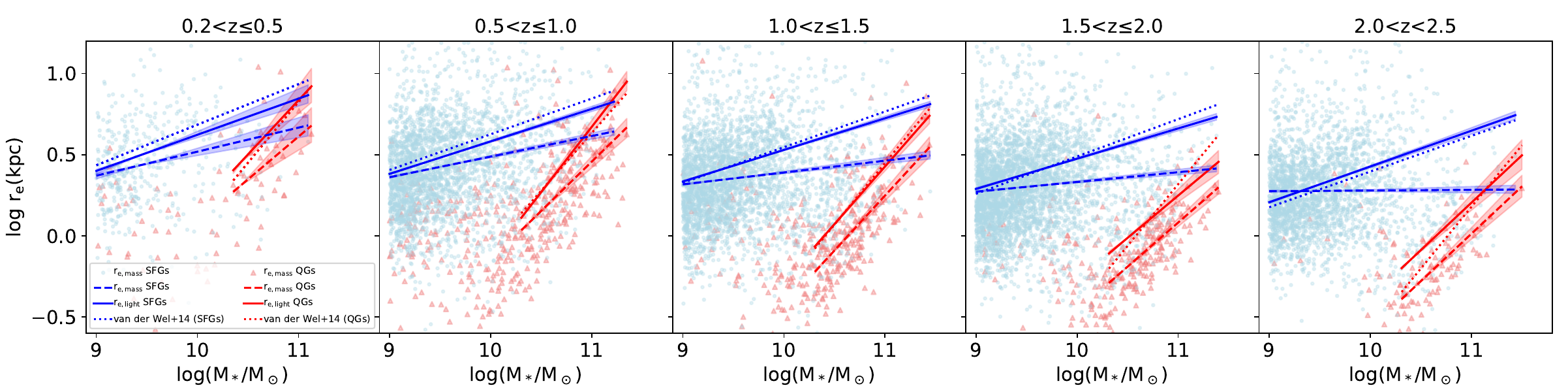}
 \caption {The mass-size relations for SFGs and QGs across five distinct redshift bins. Blue circles and red triangles denote data points of SFGs and QGs, respectively; correspondingly, the solid lines represent the best-fit lines for $r_{\rm e, light}$, and the dashed lines represent the best-fit lines for $r_{\rm e, mass}$. The best-fit results are listed in Table \ref{tab:2}. The dot-dashed lines indicate the median relations in the rest-frame optical (0.5µm) from \cite{vanderWel_2014}. The shaded regions around the fitting lines are determined by the 1$\sigma$ confidence intervals obtained from 500 bootstrap resamplings.}
\label{fig:9}
\end{figure*} 

\subsection{Difference between $r_{\rm e, mass}$ and $r_{\rm e, light}$}  \label{4.1}

As mentioned in the Introduction, many studies have shown significant differences between $r_{\rm e, mass}$ and $r_{\rm e, light}$ (e.g., \citealt{Suess_2019, Mosleh_2020, Miller_2023}). The ratio of $r_{\rm e, mass} / r_{\rm e, light}$ reflects the color gradient within a galaxy. In the absence of a color gradient, $r_{\rm e, mass}$ and $r_{\rm e, light}$ would be equal. Generally, a strong negative color gradient could result in $r_{\rm e, mass} / r_{\rm e, light} < 1$, while a positive color gradient could lead to $r_{\rm e, mass} / r_{\rm e, light} > 1$. Since a galaxy's color gradient is closely tied to its growth processes, studying the relationship between $r_{\rm e, mass} / r_{\rm e, light}$ and other physical properties can provide us with valuable insights into the mechanisms driving galaxy evolution.

In Figure \ref{fig:7}, we have presented the distribution of $r_{\rm e, mass} / r_{\rm e, light}$ as a function of galaxy's half-light radius, stellar mass, and rest-frame $U-V$ color. From top to bottom, we show the results at $0.2<z\le0.5$, $0.5< z\le 1.0$,  $1.0< z\le 1.5$,  $1.5< z\le 2.0$,  and $2.0<z<2.5$, respectively. In each panel, the red and blue points represent the distributions for QGs and SFGs, respectively. To better illustrate the trends between $r_{\rm e, mass} / r_{\rm e, light}$ and other galaxy properties, we fit a linear relationship between $\log (r_{\rm e, mass} / r_{\rm e, light})$ and other corresponding physical properties using the least-square method. The specific fitting equation and the corresponding best-fit results are presented in Table \ref{tab:1}. The $1\sigma$  errors on the fit are estimated by performing 500 bootstrap simulations. The best-fit relationships are also shown as solid lines in Figure \ref{fig:4}. Additionally, in the rightmost column, we present the distributions of $\log (r_{\rm e, mass} / r_{\rm e, light})$ for SFGs and QGs across different redshift ranges, with dashed lines indicating the corresponding medians.

As shown in Figure \ref{fig:7}, $r_{\rm e,mass}$ is systematically smaller than $r_{\rm e,light}$ for both populations across all redshifts. The median value of $\log (r_{\rm e, mass} / r_{\rm e, light})$ for SFGs is -0.07, -0.06, -0.07, -0.05, and -0.03 at $0.2<z\le0.5$, $0.5< z\le 1.0$,  $1.0< z\le 1.5$,  $1.5< z\le 2.0$,  and $2.0<z<2.5$, respectively. For QGs, the corresponding median values are -0.11, -0.14, -0.15, -0.15, and -0.16, respectively. 
This is consistent with many previous studies. Using HST observations in the CANDELS fields, \cite{Mosleh_2017} and \cite{Suess_2019} all found that $r_{\rm e,mass}$ is smaller than by approximately 30\%–45\% (about 0.15 dex) for SFGs and 30\%–50\% for QGs. Moreover, by using rest-frame near-infrared morphology as a tracer of the stellar mass distribution, several recent studies based on JWST data have also reported similar results (e.g., \citealt{Suess_2022,ji2024jadesrestframeuvtonirsize,Martorano_2024}).

From Figure~\ref{fig:7}, we also find that $r_{\rm e,mass}/r_{\rm e,light}$ is correlated with the physical properties of galaxies. Specifically, log($r_{\rm e,mass}/r_{\rm e,light}$) shows negative correlations with $r_{\rm e,light}$, stellar mass, and $U-V$ color for both SFGs and QGs. Although \citet{Szomoru_2013} found no strong correlation between $r_{\rm e,mass}/r_{\rm e,light}$ and galaxy stellar mass. More studies, however, have shown results consistent with ours. For example, \citet{Suess_2019} found that $r_{\rm e,mass}/r_{\rm e,light}$ clearly depends on stellar mass and other physical properties. Some recent JWST-based studies also find this correlation. Using JWST observations in the CEERS field, \citet{vanderWel_2024} found that galaxy color gradients correlate with both stellar mass and specific star formation rate (sSFR). By using rest-frame $1\ \mu m$ morphology as a tracer of the stellar mass distribution, \citet{Suess_2022} also find a significant correlation between $r_{\rm e,mass}/r_{\rm e,light}$ and galaxy physical properties. This may reflect that, during the process of stellar mass growth, galaxy centers become older, dustier, and more metal-rich, which in turn leads to stronger color gradients.

Additionally, We have also investigated the evolution of $r_{\rm e,mass}/r_{\rm e,light}$ with redshift. In Figure~\ref{fig:8}, we present the redshift evolution of the median ratio $r_{\rm e,mass}/r_{\rm e,light}$ for SFGs and QGs, computed in six redshift intervals: $0.2<z<0.5$, $0.5<z<0.8$, $0.8<z<1.1$, $1.1<z<1.4$, $1.4<z<1.7$, $1.7<z<2.0$, and $2.0<z<2.5$. The corresponding error bars are estimated using 500 bootstrap resamplings, and the error for each galaxy is also accounted for by adding random perturbations. As shown in the figure, both populations exhibit $r_{\rm e,mass}/r_{\rm e,light} < 1$ at all redshifts, consistent with the trends seen in Figure~\ref{fig:7}. For QGs, the ratio remains roughly constant at $z>1$ with a median value of $\sim 0.7$, and shows a slight increase as the redshift decreases from 1. For SFGs, $r_{\rm e,mass}/r_{\rm e,light}$ increases with redshift at $z>1.5$.

However, some previous studies have reported somewhat different results. For SFGs, \citet{Suess_2019} and \citet{Miller_2023} all found that $r_{\rm e,mass}/r_{\rm e,light}$ increases with redshift at $z>1$, which is consistent with our results. However, for QGs, they also found an increasing ratio toward higher redshift, which is in contrast with our result. It should be worth noting that in their work, they only used HST data, which can only cover the rest-frame optical at $z>1$, making it difficult to break the degeneracies between stellar age, metallicity, and dust. By combining HST and JWST observations, \citet{vanderWel_2024} also found that $r_{\rm e,mass}/r_{\rm e,light}$ decreases with increasing redshift, which is consistent with our results. This may indicate that JWST data can help us to derive more accurate physical properties at high redshift, or its higher sensitivity enables us to better detect the faint structures in the outskirts of galaxies.

\begin{table*}[htbp]
\centering
\caption{The best-fit parameters of the relationship between the different radii and stellar mass for SFGs and QGs, given by the fitting formula  $r_{\rm e}(M)/\text{kpc} = A \times M^\alpha$.}
\label{tab:2}
\begin{tabular}{c|c|c|c|c} 
\hline\hline
\multirow{2}{*}{$z$}& \multicolumn{2}{c}{SFGs} & \multicolumn{2}{c}{QGs}\\ 
\cline {2-3} \cline {4-5}& $\alpha \rm(r_{ e, light})$& $\alpha \rm(r_{ e, mass})$& $\alpha \rm(r_{ e, light})$& $\alpha \rm(r_{ e, mass})$\\ 
\hline
$0.2  <  z \leq 0.5$& $0.22^{+0.04}_{-0.03}$& $0.15^{+0.04}_{-0.04}$& $0.66^{+0.19}_{-0.15}$& $0.52^{+0.15}_{-0.17}$\\ 
\hline
$0.5  <  z \leq 1.0$& $0.20^{+0.01}_{-0.01}$& $0.12^{+0.01}_{-0.01}$& $0.80^{+0.06}_{-0.06}$& $0.61^{+0.06}_{-0.07}$\\ 
\hline
$1.0  <  z \leq 1.5$& $0.20^{+0.01}_{-0.01}$& $0.08^{+0.01}_{-0.01}$& $0.71^{+0.08}_{-0.05}$& $0.68^{+0.06}_{-0.06}$\\ 
\hline
$1.5  <  z \leq 2.0$& $0.19^{+0.01}_{-0.01}$& $0.06^{+0.01}_{-0.01}$& $0.56^{+0.03}_{-0.12}$& $0.54^{+0.04}_{-0.05}$\\ 
\hline
$2.0  <  z < 2.5$& $0.22^{+0.01}_{-0.01}$& $0.01^{+0.01}_{-0.01}$& $0.59^{+0.08}_{-0.09}$& $0.58^{+0.06}_{-0.06}$\\ 
\hline
\end{tabular}
\end{table*}

\subsection{The Mass-Size Relation}

Given that the ratio of $r_{\rm {e,mass}}$ to \(r_{\rm {e,light}}\) varies with stellar mass, the relationship between stellar mass and \(r_{\rm {e,mass}}\) should differ from that between stellar mass and \(r_{\rm {e,light}}\). This section presents these size-mass relations for both SFGs and QGs across a range of redshifts from 0.2 to 2.5.

Similar to previous studies (e.g., \citealt{Mowla_2019, Miller_2019}), we adopt the following functional form to fit our mass-size relation using the least-squares method:
\begin{equation}
    r_{\rm e}(M)/\text{kpc} = A \times M^\alpha,
    \label{eq:reff_mstar}
\end{equation}
where $M$ is defined as $M_\ast/(5 \times 10^{10} M_\odot)$. We fit the mass–size relation in five redshift intervals using a least‐squares approach. In order to better compare with previous work, for SFGs we include the full sample, whereas for QGs we restrict the fit to galaxies with $M> 2 \times 10^{10} M_\odot$. The best-fit results are presented in Figure \ref{fig:9}, and the corresponding slopes $\alpha$ are shown in Table \ref{tab:2}. For each fit, the errors are obtained using 500 bootstrap resamplings. 

Across all redshift intervals, we find that SFGs exhibit systematically larger sizes than QGs at fixed stellar mass, irrespective of whether $r_{\rm e,light}$ or $r_{\rm e,mass}$ is used. This trend is consistent with expectations from structural evolution, where QGs tend to be more centrally concentrated. Additionally, we find that for QGs at lower masses, both for $r_{\rm {e,light}}$ and $r_{\rm {e,mass}}$, there is a significantly flatter part of the size-mass distribution. A similar observational trend had also been presented by \cite{vanderWel_2014} using observations from HST and \cite{Martorano_2024} using observations from JWST. Notably, the slopes of the mass-size relations derived from $r_{\rm e,light}$ are steeper than those based on $r_{\rm e,mass}$ for both SFGs and QGs. This pattern holds across all redshift bins and has been found in many previous studies (e.g., \citealt{vanderWel_2014,Suess_2019,Miller_2023,Martorano_2024,ji2024jadesrestframeuvtonirsize,Jia_2024} ). The difference in slope can be understood in light of our earlier finding that massive galaxies tend to exhibit smaller $r_{\rm e,mass} / r_{\rm e,light}$ ratios, reflecting stronger negative color gradients. This implies that when measuring galaxy structure based solely on luminosity distribution, the increase in size attributable to stellar mass growth is overestimated.

Both the $r_{\rm e,mass}$-mass and $r_{\rm e,light}$-mass relations have been extensively studied in the literature (e.g., \citealt{vanderWel_2014,Mosleh_2017,Mowla_2019,Suess_2019,Miller_2023,Martorano_2024,ji2024jadesrestframeuvtonirsize,Jia_2024} ). One of the most well-known results is from \cite{vanderWel_2014}, who derived the $r_{\rm e,light}$-mass relation using data from the CANDELS fields. In Figure \ref{fig:9}, we also present their relation as the dot-dashed lines, which show excellent agreement 
between our measured $r_{\rm e,light}$ slopes and their reference values within mutual uncertainties. This consistency further supports the reliability of our results. When examining the $r_{\rm e,mass}$-mass relation, we find that for SFGs, the relation becomes shallower at higher redshifts, with the slope approaching zero at $z\sim 2.5$. Similar trends have been reported in previous studies. For example, \cite{Miller_2023} found a slope of $\alpha = 0.17$ at $z\sim 1.1$ and $\alpha=0.01$ at $z\sim 1.9$. \cite{Martorano_2024} also found that the $r_{\rm e,mass}$-mass relation becomes flatter at higher redshift, although in their study $\alpha$ remains above 0.05 even at $z>2$. Nonetheless, these results provide a valuable starting point. If the observed trend of decreasing slope with increasing redshift continues, it is possible that the relation may become flat—or even inverted—at earlier epochs, a scenario that has been supported by both observational \citep{Ormerod_2024} and simulation-based studies (e.g., \citealt{Genel_2017,Costantin_2023}). As for QGs, the slope of their $r_{\rm e,mass}$-mass relation shows little evolution with redshift, consistent with findings from several other studies (e.g., \citealt{ji2024jadesrestframeuvtonirsize,vanderWel_2024}). This may indicate that QGs grow through a similar mechanism across different cosmic epochs.

\begin{figure*}
 \includegraphics[width=1\linewidth]{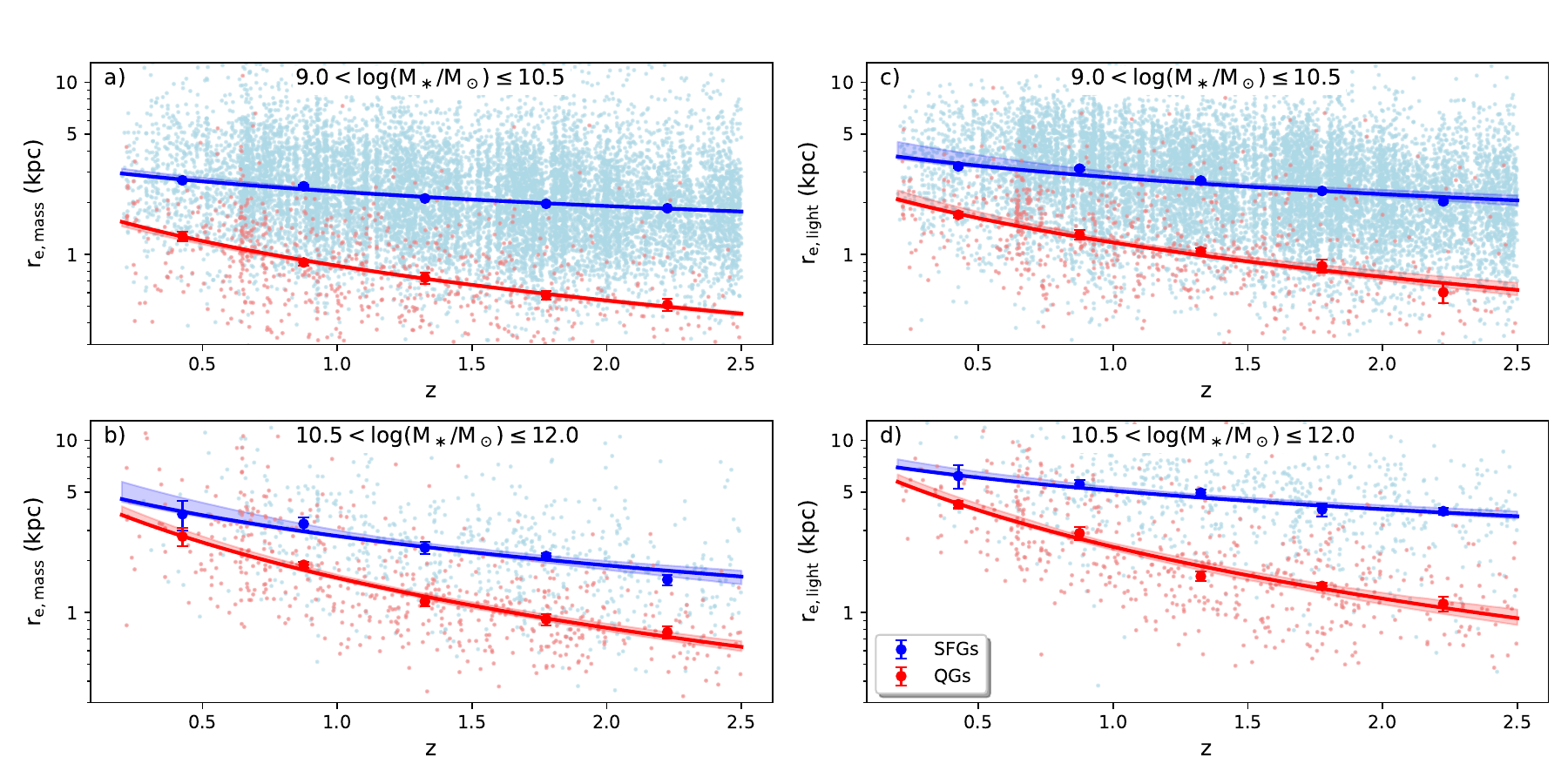}
 \caption {Evolutions of $r_{\rm e,mass}$ (left panels) and $r_{\rm e,light}$ (right panels) for SFGs (blue) and QGs (red) at $0.2<z<2.5$ in $10^{9.0}M_\odot<M_*\le10^{10.5}M_\odot$ (top panels) and $10^{10.5}M_\odot <M_*\le 10^{12.0}M_\odot$ (bottom panels), respectively. The light dots represent the results for individual galaxies. The blue and red points represent the median values for SFGs and QGs, respectively, with error bars indicating 1$\sigma$ confidence intervals derived from 500 bootstrap resamples. The lines represent the best fits to the median values, assuming a relationship of (\(1 + z)^\gamma\).} 
\label{fig:10}
\end{figure*}   

\subsection{Size Evolution at Fixed Stellar Mass} 
Given the influence of redshift on the ratio of $r_{\rm {e,mass}}$ to $r_{\rm {e,light}}$, the evolution of $r_{\rm e,mass}$ could differ from $r_{\rm e,light}$. In this section, we examine the evolution of galaxy sizes within the redshift range $0.2 < z < 2.5$. To minimize the effect that galaxy's size is closely related to stellar mass, we divide our sample into two stellar mass bins ($10^{9.0}M_{\odot}<M_*\le 10^{10.5}M_{\odot}$ and $10^{10.5}M_{\odot}<M_*\le 10^{12.0}M_{\odot}$) for further analysis. 

The distributions of galaxy size as a function of redshift are shown in Figure \ref{fig:10}. The evolution of $r_{\rm e,mass}$ is shown in left panels a)-b), while the results for $r_{\rm e,light}$ are presented in right panels c)-d). In each panel, the red and blue points represent the median size within different redshift intervals of SFGs and QGs, respectively. The evolution of galaxy sizes is described by the relation $r_e \propto (1 + z)^\gamma$. We fit all the median points in Figure \ref{fig:10}. The fitting method and the error estimation method remain consistent with those in the analysis of the mass-size relation. The best fitting results are also plotted as solid lines in this figure, while the best-fit parameters $\gamma$ are listed in Table \ref{tab:3}. When we directly fit the individual points instead of the medians, the results remain unchanged.

In Figure~\ref{fig:10}, it can be observed that the size evolution of QGs is significantly faster than that of SFGs, regardless of whether considering $r_{\rm {e,mass}}$ and $r_{\rm e,light}$. This observation may suggest distinct mass assembly mechanisms for QGs and SFGs. When examining the size evolution of galaxies with different stellar masses, the slope for massive galaxies appears slightly steeper than that for less massive galaxies, although the difference is not pronounced. This trend has also been demonstrated in other studies. For example, by selecting galaxies in the CANDELS fields, \cite{vanderWel_2014} found a slope of $\gamma= -1.01$ for early-type galaxies and $\gamma= -0.52$ for late-type galaxies at a stellar mass of $10^{10.25} M_{\odot}$. In contrast, for galaxies with a stellar mass of $10^{11.25}M_{\odot}$, the slope becomes steeper, with $\gamma = -1.32 $ for early-type galaxies and -0.80 for late-type galaxies. This dependency between galaxy size evolution and stellar mass could be attributed to massive galaxies undergoing more frequent minor mergers, which drive faster size evolution.

\begin{table}[ht]
\centering 
\caption{The best-fitting parameter $\gamma$ of different radii ($r_{\rm e} \propto (1 + z)^{\gamma}$, with $0.2<z < 2.5$) for SFGs and QGs in different stellar mass bins.}\label{tab:3}
\begin{tabular}{c|c|c|c}
\hline\hline
 Radius & $\log M_*/M_\odot$ & SFGs  & QGs \\
\hline
 \multirow{2}{*}{$r_{\rm e, light}$} & 9.0-10.5 & $-0.55^{+0.06}_{-0.25}$& $-1.04^{+0.07}_{-0.05}$\\
 \cline{2-4}
  & 10.5-12.0 & $-0.60^{+0.06}_{-0.11}$& $-1.72^{+0.10}_{-0.20}$\\
\hline
 \multirow{2}{*}{$r_{\rm e, mass}$} & 9.0-10.5 & $-0.47^{+0.03}_{-0.08}$& $-1.10^{+0.08}_{-0.03}$\\
 \cline{2-4}
  & 10.5-12.0 & $-0.97^{+0.11}_{-0.30}$& $-1.64^{+0.08}_{-0.14}$\\
\hline
\end{tabular}
\end{table}

In this study, we find clear size evolution for both SFGs and QGs, regardless of the radius definition adopted. Specifically, over the redshift interval $z=$1.25-2.25, massive SFGs in our sample grow by $0.65^{+0.05}_{-0.20}$ kpc, while QGs grow by $0.51^{+0.01}_{-0.05}$ kpc. This growth is significantly larger than that reported in previous studies based on HST data (e.g., \citealt{Suess_2019, Miller_2023}), which found almost negligible size evolution over the same redshift range. 

In several recent studies based on JWST data, it has also been found that galaxy sizes evolve significantly with redshift, regardless of the definition of radius used (e.g., \citealt{vanderWel_2024,Clausen_2025,ji2024jadesrestframeuvtonirsize,Jia_2024,Martorano_2024}). With a sample of 161 QGs from the JADES field, \cite{ji2024jadesrestframeuvtonirsize} found that the size evolution rate at rest-frame 1 $\mu m$ ($\gamma = -1.15^{+0.25}_{-0.22}$) is a little faster than that at rest-frame 0.5 $\mu m$ ($\gamma = -1.04^{+0.25}_{-0.26}$). Similarly, in the work of \cite{vanderWel_2024}, they found that the size evolution for massive QGs can also be significant ($\gamma = -1.64$ for $r_{\rm e, light}$ and $\gamma = -1.72$ for $r_{\rm e, mass}$). For SFGs, \cite{Martorano_2024} used data from the COSMOS-Web and PRIMER-COSMOS fields and found that the size–redshift relation slope $\gamma$ ranges from –0.74 to –1.16 when using galaxy sizes measured at rest-frame 1.5 $\mu$m. Similarly, \cite{Jia_2024} reported $\gamma=-1.08$ for galaxy sizes measured at rest-frame 1 $\mu$m, and $\gamma=-1.04$ at 0.45 $\mu$m. 

Although the inferred rates of evolution differ somewhat across studies due to variations in sample selection and methodology, they consistently show that galaxy sizes increase over time—regardless of whether the sizes are luminosity-weighted or mass-weighted. Additionally, some simulations also support this finding. Using the EAGLE simulation, \cite{Furlong_2016} found that the mean size of galaxies increases by 50\% to 75\% from $z=2$ to $z=1$. The similar results had also been seen in TNG-100 by \cite{Genel_2017}. These results indicate that galaxy evolution is accompanied by an increase in size and highlight the importance of JWST data for studying this process in detail.

\section{Discussion } \label{sec:5}
\subsection{Comparison with 1D results} \label{5.1}

\begin{figure*}
 \includegraphics[width=1\linewidth]{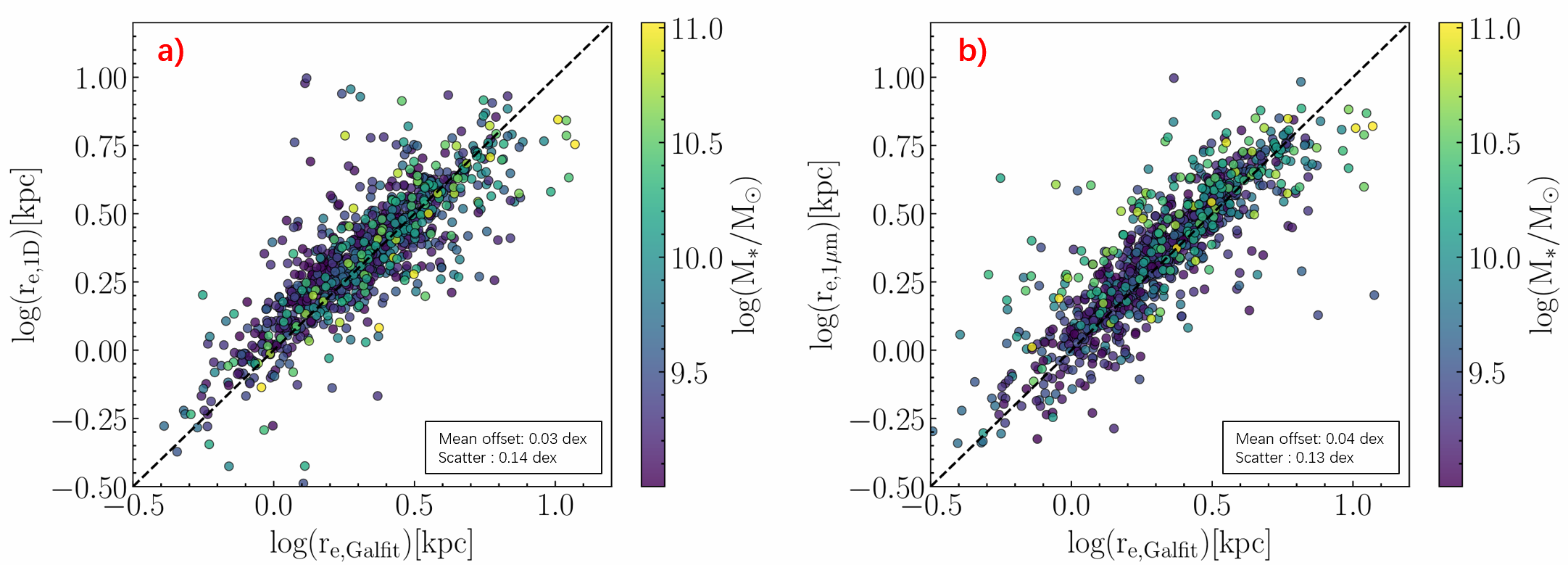}
 \caption{Panel (a): Comparison between $r_{\rm e,mass}$ derived from \texttt{GALFIT} 2D fitting ($r_{\rm e,Galfit}$) and from 1D stellar mass surface density profile fitting ($r_{\rm e, 1D}$). The two measurements show good agreement. Panel (b): Comparison between $r_{\rm e,Galfit}$ and the effective radius measured at the rest-frame $1\mu m$.} 
\label{fig:14}
\end{figure*}  

In this work, our main conclusions are based on $r_{\rm e,Galfit}$, the effective radius derived from 2D stellar mass maps using \texttt{GALFIT} (Method 1; Section \ref{sec:3.3.1}). To verify the reliability of this method, we also derive $r_{\rm e,1D}$ from 1D $\Sigma_\ast$ profiles (Method 2; Section \ref{sec:3.3.2}) and compare the results for galaxies in the JADES-GDS field, as shown in panel a) of Figure \ref{fig:14}. The two measurements show good consistency, with a median difference of only 0.03 dex and a scatter of approximately 0.14 dex. Furthermore, since galaxy morphologies observed in the rest-frame near-infrared are good tracers of the stellar mass–weighted structure (e.g., \citealt{vanderWel_2024,ji2024jadesrestframeuvtonirsize}), we compare $r_{\rm e,Galfit}$ with the effective radius measured at rest-frame $1 \mu$m in panel b). This comparison also shows good agreement, with a median difference of 0.04 dex and a median scatter of 0.13 dex. To account for potential variations in observational depth across different fields, we repeat the analysis using data from the PRIMER-COSMOS field and find consistent results. These comparisons demonstrate the robustness of our methods across different fields and measurement approaches.

Nevertheless, we need to reminder that our analysis does not explicitly account for non-S\'{e}rsic structures in galaxies, which may bias structural parameters derived from single-S\'{e}rsic fits. Although our one-dimensional method likely mitigates these effects to some extent, a thorough assessment of non-S\'{e}rsic structures may still be needed. However, such a study lies beyond the scope of this study and plan to explore this issue in greater detail in future work.

\begin{figure*}
 \includegraphics[width=1\linewidth]{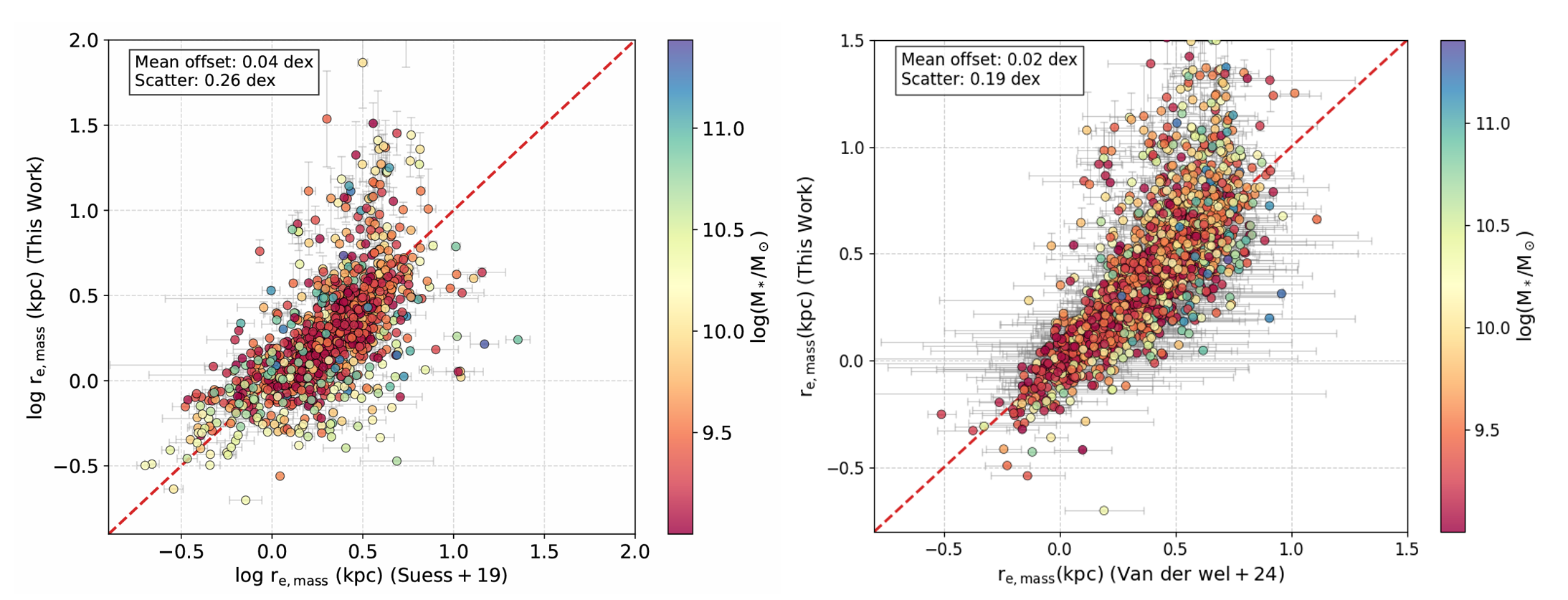}
\caption{Comparison of the measured $r_{\rm e, mass}$ values in this study with those reported by \cite{Suess_2019} and \cite{vanderWel_2024}. The scatter uncertainties are set according to the measurement errors. Relative to \cite{Suess_2019}, the results show an average offset of 0.04 dex with a scatter of 0.26 dex, while the comparison with \cite{vanderWel_2024} yields a smaller average offset (0.02 dex) and scatter (0.19 dex).}
\label{fig:13}
\end{figure*}

\subsection{Comparison with Previous Studies} 
Building on the internal consistency of our methods, we now situate our results within the context of existing literature. Several previous studies have provided catalogs of measured $r_{\rm e,mass}$ values (e.g., \citealt{Szomoru_2013, Suess_2019, Mosleh_2020, Miller_2023, vanderWel_2024}). To further assess the reliability of our method, we compare our results with those from earlier works, focusing on \cite{Suess_2019} and \cite{vanderWel_2024}. The former is widely used as a benchmark, while the latter is likewise based on JWST observations, as in our study. In the case of \cite{Suess_2019}, multiple approaches were used to derive $r_{\rm e,mass}$, and here we adopt the results from Method 1, which the authors recommend as their preferred strategy. For \cite{vanderWel_2024}, we specifically compare with the results obtained from their 2D measurement method.

As shown in Figure \ref{fig:13}, we match the results of this study with those reported by \cite{Suess_2019} and \cite{vanderWel_2024}. Compared with \cite{Suess_2019}, the overall offset between the two measurements is small (0.04 dex), but the scatter is more pronounced at about 0.26 dex. This finding is consistent with previous comparative analyses: \cite{Mosleh_2020} reported an average offset of 0.04 dex and a scatter of $\sim$0.20 dex when comparing their results with those of \cite{Suess_2019}, while \cite{Miller_2023}, using a Bayesian multi-Gaussian expansion technique to model galaxy profiles and measure $r_{\rm e,mass}$, found a mean offset of 0.05 dex and a scatter of 0.20 dex relative to \cite{Suess_2019}. Notably, despite differences in both data and methodology, both studies similarly indicated that the \cite{Suess_2019} measurements tend to be systematically larger. In contrast, our comparison with \cite{vanderWel_2024} yields a smaller overall offset (0.02 dex) and scatter (0.19 dex). Relative to \cite{Suess_2019}, the better agreement with \cite{vanderWel_2024} suggests that JWST data are likely critical for obtaining robust estimates of $r_{\mathrm{e,mass}}$. This may be because incorporating JWST data allows SED fitting to better disentangle the degeneracy between dust attenuation and stellar population age. Taken together, the consistent offsets and scatters across different studies demonstrate the robustness of our results.

\subsection{Mock recovery test} \label{5.3}

\begin{figure*}
 \includegraphics[width=1\linewidth]{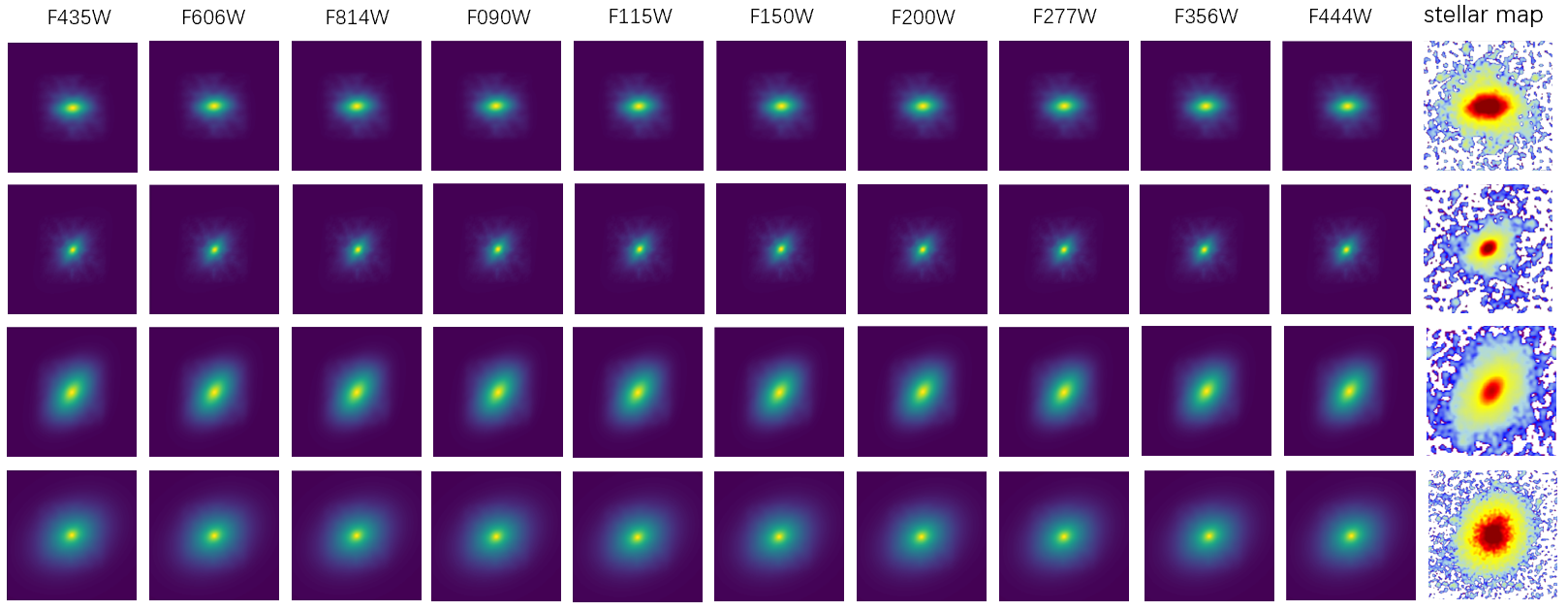}
 \caption{Examples of mock galaxies showing multi-band simulated images (left) and corresponding stellar mass maps (rightmost column). All filters adopt identical structural parameters ($n$, $r_{50}$, $b/a$, and PA) and are convolved with the F444W-band PSF.The stellar mass maps are derived from pixel-by-pixel SED fitting.} 
\label{fig:11}
\end{figure*}

\begin{figure}
 \includegraphics[width=1\columnwidth]{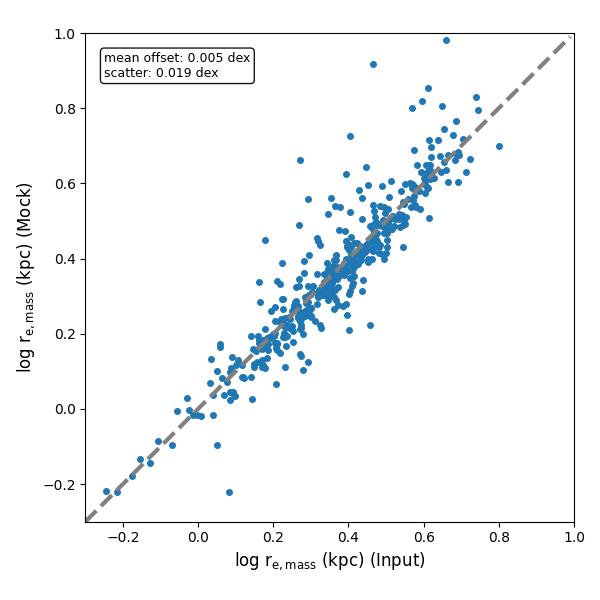}
 \caption{The simulated $r_{\rm e, mass}$ was compared with the input radius, revealing an average deviation of less than 0.001 dex, demonstrating excellent consistency.} 
\label{fig:12}
\end{figure}

To further validate the reliability of our method, we perform mock recovery tests using simulated galaxies. Following \cite{Mosleh_2020}, we randomly select 500 real galaxies from our catalog to construct single-S\'{e}rsic models for each filter, using identical structural parameters ($n$, $r_{50}$, $b/a$, and PA). The range of these parameters is chosen to match the observed distribution of galaxies in the stellar mass–size plane (see Figure \ref{fig:9}). All models are convolved with the F444W-band PSF. To remove any inherent color gradients, we assume the same structural shape in all filters, so the simulated galaxies have identical shapes at different wavelengths. The total flux of each model in each filter is scaled to match the flux of the corresponding real galaxy, ensuring that the total stellar mass is conserved. Finally, the models are inserted into empty regions of the mosaic images that contain realistic background noise, to generate mock observations. Figure \ref{fig:11} shows four randomly selected examples of these multi-band mock galaxies generated using GALFIT.

We then recover the stellar mass maps of these mock galaxies using the same pixel-by-pixel SED fitting procedure described in Section~\ref{sec:3.3.1}. The rightmost column of Figure \ref{fig:11} shows examples of the recovered stellar mass maps. We measure $r_{\rm e, mass}$ for each mock galaxy using \texttt{GALFIT} and compare the results with the input effective radii. As shown in Figure \ref{fig:12}, the recovered $r_{\rm e, mass}$ values are in excellent agreement with the input values, with a median difference of 0.005~dex and a scatter of 0.019~dex. This demonstrates that our pixel-by-pixel SED fitting method reliably recovers the stellar mass distribution of galaxies.

\subsection{Possible physical explanation}

In this study, we utilize JWST observations within the CANDELS fields to systematically examine the differences between $r_{\rm e, mass}$ and $r_{\rm e, light}$, as well as to investigate how the choice of radius definition influences the size--mass relation and its redshift evolution. Overall, our observations support the inside-out growth scenario, in which the central regions of galaxies form and evolve earlier than their outskirts, providing key insights into the structural development of galaxies.

In Figure \ref{fig:7}, we have found a negative correlation between $r_{\rm e, mass}/r_{\rm e, light}$ and galaxy stellar mass. This trend can be naturally interpreted in the framework of inside-out growth: as the stellar mass of galaxies increases, their centers become older and contain more dust, while the outskirts continue forming stars, which may lead to a steeper negative color gradient (e.g., \citealt{Tacchella_2016,Jin_2024,Abdurro'uf_2023}).

Moreover, we observe distinct redshift evolution of $r_{\rm e, mass}/r_{\rm e, light}$ between SFGs and QGs. At $z>1.5$, $r_{\rm e, mass}/r_{\rm e, light}$ of SFGs increase with redshift. This may be because, at high redshift, SFGs may maintain active star formation across all radii, resulting in relatively flat color gradients (e.g., \citealt{Genel_2017,Costantin_2023,Ormerod_2024}). This is also consistent with our finding that SFGs at high redshift exhibit a shallower slope in the size–mass relation. At high redshift, star formation occurs across all radii of galaxies, which increases their stellar mass but may not significantly change their size. In contrast, at low redshift, star formation is primarily concentrated in the outskirts of galaxies, so the increase in stellar mass leads to a more pronounced growth in galaxy size. While for QGs, their central regions are already older and richer in metals or dust; over time, outer regions gradually accumulate metals and dust or undergo mergers that redistribute stellar populations, flattening the color gradients and stabilizing or slightly increasing $r_{\rm e, mass}/r_{\rm e, light}$ at lower redshifts (e.g., \citealt{Naab_2009,Oser_2012,Whitaker_2013,Pacifici_2016}).

\section{Summary} \label{sec:6}
In this study, by combining HST and JWST data, we construct stellar mass maps for galaxies with redshifts in the range of $0.2<z<2.5$ using pixel-by-pixel SED fitting. Then we measure their $r_{\rm {e,mass}}$ and $r_{\rm {e,light}}$ for these galaxies to investigate their mass-size relation and size evolution. Our conclusions are summarized below:

(1) Both QGs and SFGs exhibit $r_{\rm e,mass}$ smaller than their $r_{\rm e,light}$. The differences between these two radii are clearly negatively correlated with stellar mass, $(U - V){_{\rm rest}}$ color, and $r_{\rm e,light}$.

(2) The $r_{\rm e,mass}$ also exhibits a clear positive correlation with stellar mass. But the slope for $r_{\rm e,mass}$ is flatter than that of $r_{\rm e,light}$.

(3) Similar to $r_{\rm e,light}$, $r_{\rm e,mass}$ also exhibits clear evolution with redshift. In our sample, at fixed stellar mass, QGs increase their effective radius by a factor of $\sim3$--5 from $z \sim 2.5$ to $z \sim 0.2$, while SFGs grow more moderately, by a factor of $\sim2$. 

(4) For massive galaxies, we find $\gamma = -0.97^{+0.11}_{-0.30}$ for SFGs and $\gamma = -1.64^{+0.08}_{-0.14}$ for QGs. These trends are consistent with those observed for $r_{\rm e,light}$, indicating that both mass-weighted and light-weighted sizes follow a similar redshift evolution. 

However, in this study, we are unable to determine the physical origin of the difference between $r_{\rm e,mass}$ and $r_{\rm e,light}$ because we cannot distinguish the effects of age, metallicity, dust, and some other physical factors. Additionally, due to the wavelength limitations of NIRCam, we are unable to extend this study to galaxies at higher redshifts. In the future,  we will incorporate MIRI data to investigate $r_{\rm e,mass}$ of galaxies at even higher redshifts. We also plan to use some integral field spectroscopic data to investigate the physical origins of color gradients and explore how these gradients and $r_{\rm e,mass}$ evolve as galaxies transition from star-forming to quiescent states.

\begin{acknowledgments}
This work is supported by the National Science Foundation of China (NSFC, Grant Nos. 12233008, 12573012), the National Key R\&D Program of China (2023YFA1608100), the Strategic Priority Research Program of the Chinese Academy of Sciences (Grant No. XDB0550200), the Cyrus Chun Ying Tang Foundations, the 111 Project for ``Observational and Theoretical Research on Dark Matter and Dark Energy'' (B23042), and the China Manned Space Project. S.L. acknowledges the support from the Key Laboratory of Modern Astronomy and Astrophysics (Nanjing University) by the Ministry of Education. The numerical calculations in this paper were performed on the computing facilities of the High Performance Computing Platform at Anqing Normal University.
\end{acknowledgments}

\bibliography{ref}{}
\bibliographystyle{aasjournal}
\end{document}